%% file: JMullaney2010.tex
\documentclass[useAMS,usenatbib]{mn2e}
\usepackage{times,mathptm}
\usepackage{lscape}

\usepackage{natbib}
\usepackage{graphicx}
\usepackage{floatflt}
\usepackage{amssymb}

\newcommand{\cs}{$\textnormal{cm}^{-2}$}

\newcommand{\ergs}{ergs~s$^{-1}$}
\newcommand{\lsun}{$L_\odot$}

\newcommand{\kev}{keV}
\newcommand{\mum}{${\rm \umu m}$}

\newcommand{\nev}{$[$Ne~{\sc v}$]$}
\newcommand{\neii}{$[$Ne S3.2.2~{\sc ii}$]$}

\newcommand{\ewpah}{EW~PAH$_{\lambda11.25}$}

\newcommand{\e}[1]{$\times 10^{#1}$}

\newcommand{\fig}[1]{fig. \ref{#1}}
\newcommand{\tab}[1]{table \ref{#1}}

\newcommand{\Nh}{$N_{\rm H}$}
\newcommand{\lx}{$L_{\rm 2-10keV}$}
\newcommand{\lir}{$L_{\rm IR}$}
\newcommand{\liragn}{$L_{\rm IR}^{\rm AGN}$}
\newcommand{\ltwelve}{$\nu L_{\rm \nu}(12~\umu {\rm m})$}
\newcommand{\comment}[1]{}
\newcommand{\dcmb}{{\sc decompir}}

\newcommand{\Spitzer}{\textit{Spitzer}}
\newcommand{\Herschel}{\textit{Herschel}}
\newcommand{\IRAS}{\textit{IRAS}}
\newcommand{\Chandra}{\textit{Chandra}}

\newcommand{\Swift}{\textit{Swift}}

\title[The Infrared SEDs of AGNs]{Defining the intrinsic AGN infrared
  spectral energy distribution and measuring its contribution to the
  infrared output of composite galaxies\thanks{The intrinsic AGN
    infrared SEDs, host-galaxy templates and a procedure used to
    combine them to fit infrared photometry is available at
    http://sites.google.com/site/decompir}} \author[J.R. Mullaney et
al]{J. R. Mullaney$^{1,2}$\thanks{E-mail: james.mullaney@cea.fr},
  D. M. Alexander$^{1}$, A. D. Goulding$^{1,3}$ and
  R. C. Hickox$^{1}$\\
  $^{1}$Department of Physics, Durham University, South
  Road, Durham, DH1 3LE, U.K.\\
  $^{2}$Laboratoire AIM-Paris-Saclay, CEA/DSM/Irfu - CNRS,
  Universit\'{e} Paris Diderot, CE-Saclay,
  pt courrier 131, 91191 Gif-sur-Yvette, France\\
  $^{3}$Harvard-Smithsonian Center for Astrophysics, 60 Garden Street,
  Cambridge, MA 02138, U.S.}

\begin{document}

\date{Date Accepted}

\pagerange{\pageref{firstpage}--\pageref{lastpage}} \pubyear{2010}

\maketitle

\label{firstpage}

\begin{abstract}
  We use infrared spectroscopy and photometry to empirically define
  the intrinsic, thermal infrared spectral energy distribution (i.e.,
  6-100~\micron\ SED) of typical active galactic nuclei (i.e.,
  2-10~keV luminosity, \lx$\sim10^{42}-10^{44}$~\ergs AGNs).  On
  average, the infrared SED of typical AGNs is best described as a
  broken power-law at $\lesssim40$~\mum\ that falls steeply at
  $\gtrsim$40~\micron\ (i.e., at far-infrared wavelengths).  Despite
  this fall-off at long wavelengths, at least 3 of the 11 AGNs in our
  sample have observed SEDs that are AGN-dominated even at 60~\mum,
  demonstrating the importance of accounting for possible AGN
  contribution even at far-infrared wavelengths.  Our results also suggest
  that the average intrinsic AGN 6-100~\mum\ SED gets bluer with
  increasing X-ray luminosity, a trend seen both within our sample and
  also when we compare against the intrinsic SEDs of more luminous
  quasars (i.e., \lx$\gtrsim10^{44}$~\ergs).  We compare our intrinsic
  AGN SEDs with predictions from dusty torus models and find they are
  more closely matched by clumpy, rather than continuous, torus
  models.  Next, we use our intrinsic AGN SEDs to define a set of
  correction factors to convert either monochromatic infrared or X-ray
  luminosities into total intrinsic AGN infrared (i.e.,
  8-1000~\micron) luminosities.  Finally, we outline a procedure that
  uses our newly defined intrinsic AGN infrared SEDs, in conjunction
  with a selection of host-galaxy templates, to fit the infrared
  photometry of composite galaxies and measure the AGN contribution to
  their total infrared output.  We verify the accuracy of our SED
  fitting procedure by comparing our results to two independent
  measures of AGN contribution: (1) 12~\micron\ luminosities obtained
  from high-spatial resolution observations of nearby galaxies and (2)
  the equivalent width of the 11.25~\micron\ PAH feature.  Our SED
  fitting procedure opens up the possibility of measuring the
  intrinsic AGN luminosities of large numbers of galaxies with
  well-sampled infrared data (e.g., {\it IRAS}, {\it ISO}, {\it
    Spitzer} and {\it Herschel}).

\end{abstract}


\section{Introduction}
\label{Introduction}

The spectral energy distribution (hereafter, SED) of a continuum
source is the description of its energy output as a function of photon
frequency or wavelength.  As such, it is one of the most important
measurables in astronomy, providing information on both the physical
nature of a continuum source (e.g., stars, galaxies, active galactic
nuclei, heated dust), its influence on the surrounding matter (i.e.,
heating, ionisation state) and, when integrated over all wavelengths,
its bolometric luminosity (i.e., power output).  However, deriving an
SED is a challenging task, requiring multiple observations across the
whole of the electromagnetic spectrum and, in the case of active
galactic nuclei (hereafter, AGNs), one that is further complicated by
contamination from the ever-present host galaxy.

The level of host-galaxy contamination to the observed AGN SED is
typically highest at infrared wavelengths (i.e., 8-1000~\mum) where
the strongly rising host-galaxy SED typically dominates
(e.g. \citealt{Elvis94, Richards06, Netzer07}).  Indeed, this
contamination is so severe that the intrinsic AGN SED at these
wavelengths remains largely unconstrained by observations (except for
rare quasar-luminosity AGNs; see later).  The significance of the
uncertainties surrounding the intrinsic AGN infrared SED is clear when
we consider that the infrared portion of the average observed (i.e.,
AGN $+$ host) AGN SED represents $\gtrsim$50 per cent of the total
radio -- X-ray output (e.g., \citealt{Elvis94}).  Physical models of
the infrared-emitting dust surrounding the AGN can provide some
insights (e.g., \citealt{Siebenmorgen92, Fritz06, Schartmann08}),
although they tend to predict a very broad range of infrared AGN SEDs
as their input parameters are typically poorly constrained (see our
\S\ref{Comparison:Torus}); a situation that will be helped with a
better understanding of the true range of intrinsic AGN infrared SEDs.

Well-defined intrinsic AGN infrared SEDs may also prove a useful tool
in the measurements of AGN and star-formation activity that are
crucial to our understanding of the interactions between these two
processes that are believed to exist (e.g., \citealt{Magorrian98,
  Ferrarese00, Gebhardt00, Croton06, Bower06, Hopkins06, Hopkins07}).
It has already been demonstrated that the intrinsic mid-infrared
emission of AGNs is closely linked to the total AGN luminosity (e.g.,
\citealt{Horst08, Gandhi09}), while the far-infrared emission of a
pure star-forming galaxy provides a proxy measure of its star
formation rate (e.g., \citealt{Kennicutt98}).  Separating the total
infrared SED of a composite (i.e., AGN + star-forming) galaxy into AGN
and host components would therefore provide a measure of the levels of
AGN and star-formation activity that is largely unaffected by
obscuration or absorption; similar, in principle, to the infrared
spectral decompositions carried out by, for example,
\cite{Laurent00},\cite{Tran01} and \cite{Lutz04}.  The benefit of
using the whole infrared SED is that this technique is not restricted
to the small fraction of galaxies for which infrared spectra are
available.  Furthermore, such techniques are likely to become
increasingly popular with the availability of infrared photometry
measurements from the Spitzer and Herschel telescopes.  However, most
studies that have used SED decomposition approaches have had to rely
on either observed AGN SEDs that likely include considerable amounts
of host-galaxy contamination at infrared wavelengths or predictions
from radiative transfer models which are often poorly constrained
(e.g. \citealt{Polletta07, Fiore08, Fiore09,
  Pozzi10,Hatziminaoglou10}).  Of course, measurements derived from
decomposing the infrared SED should be regarded as complementary to
other measures of AGN activity that rely on other observables (e.g.,
X-ray luminosity, optical and infrared spectroscopy, mid-infrared
photometry etc.; \citealt{Kauffmann03, Heckman04, Hickox07, Hickox09,
  Goulding09}).

In their \citeyear{Netzer07} paper, \citeauthor{Netzer07} derived the
average intrinsic infrared SED of a sample of luminous PG quasars by
subtracting the host-galaxy component from the average observed SED,
showing that their average intrinsic AGN SED falls rapidly longward of
$\sim20$~\micron.  However, such luminous quasars are rare among the
AGN population and it is not clear whether the intrinsic quasar SED
can be applied to the much larger, less luminous (i.e., $L_{\rm
  2-10keV}<10^{44}$~\ergs) population of AGNs which represents the
majority of the integrated accretion output of AGNs in the Universe
(e.g., \citealt{Ueda03, Barger05, Hasinger05, Aird10}).  In this paper
we extend the work of \cite{Netzer07} to include more typical, lower
luminosity AGNs and demonstrate that the intrinsic AGN infrared SED
appears to be linked to AGN luminosity.  To do this, we first identify
a sample of AGNs that show minimal amounts of host-galaxy
contamination at MIR wavelengths, then carefully subtract any
host-galaxy component that can still dominate at far-infrared
wavelengths (see \S\ref{Sample}, \S\ref{Templates:SB}). We compare the
derived intrinsic AGN SEDs against other, more commonly assumed, AGN
SEDs including those derived from quasars and radiative transfer
models in \S\ref{Comparison}.  In \S\ref{Tests} we introduce a
procedure that uses these intrinsic AGN SEDs, together with a sample
host-galaxy templates, to measure the AGN and host-galaxy
contributions to the infrared output of a sample of local composite
galaxies and demonstrate the obtained values agree with results from
other, independent measures.  In \S\ref{Application} we outline two
obvious applications of these analyses: (1) defining correction
factors to convert 12~\micron\ and X-ray luminosities to total
intrinsic AGN infrared luminosities and (2) measuring the intrinsic
AGN power using infrared photometry alone. We summarise our findings
in \S\ref{Summary}.

Throughout this work we adopt $H_{0}=71$~km~s$^{-1}$~Mpc$^{-1}$,
$\Omega_{\rm M}=0.27$, and $\Omega_{\Lambda}=0.73$.

\section{Sample Description}
\label{Sample}

\begin{figure}
\includegraphics[width=80mm]{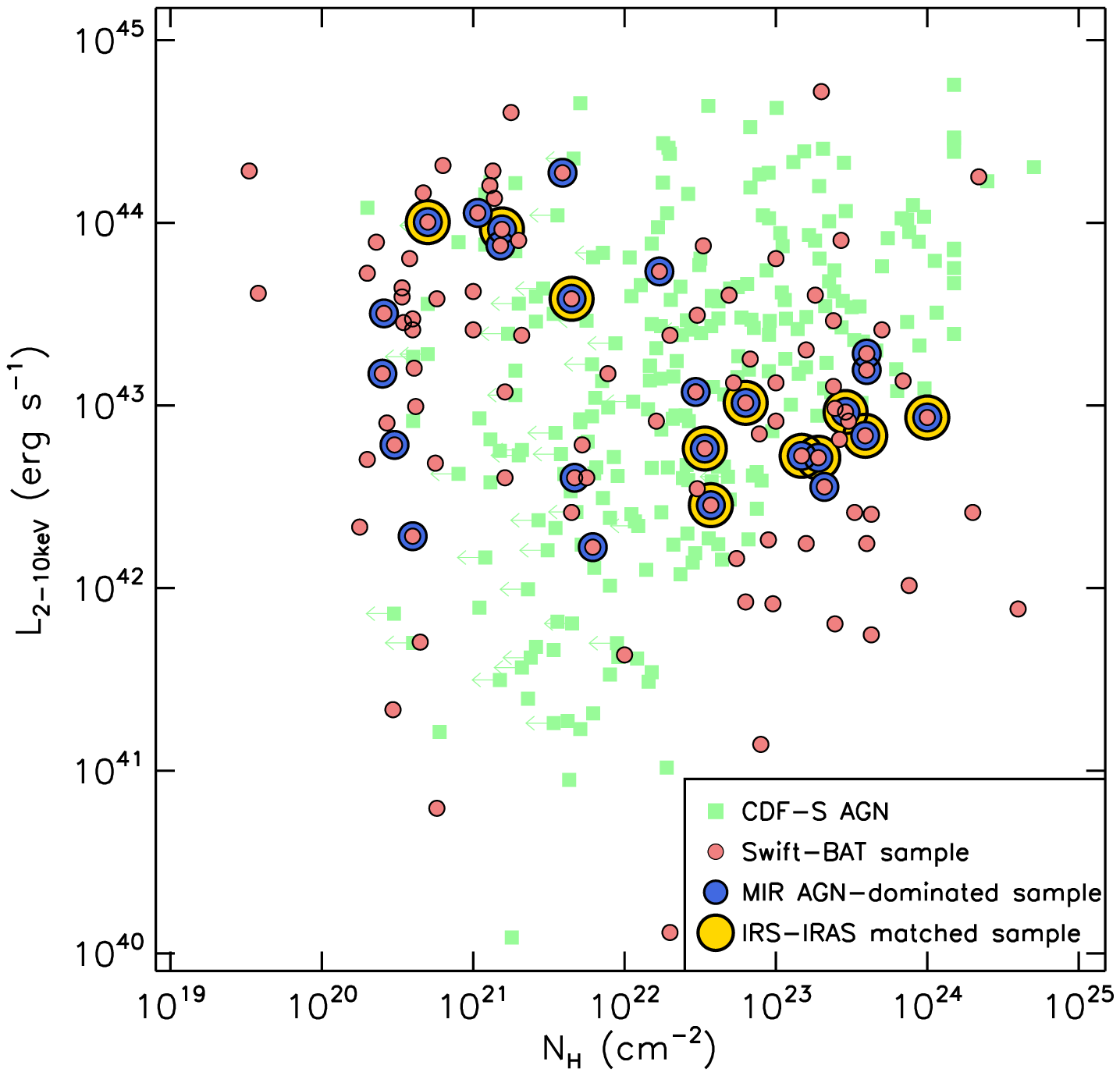}
\caption{The distribution of intrinsic 2-10~keV luminosities (i.e.,
  \lx) and hydrogen column densities (i.e., \Nh) of the \Swift-BAT
  sample of AGNs which we use to define the intrinsic AGN infrared SED
  (circles; taken from \protect \citealt{Winter09}).  In the
  background (small squares) we also show the \lx--\Nh\ distribution
  of galaxies in the \Chandra\ {\it Deep Field-South} (i.e., CDF-S;
  \protect \citealt{Tozzi06}).  The intrinsic properties of these
  local (i.e., \Swift-BAT) and high-redshift (i.e., CDF-S) samples of
  galaxies are well matched.  We highlight those 25 and 11 AGNs (see
  key) that we use to define the range of intrinsic AGN mid-infrared
  and mid to far-infrared SEDs. These subsamples cover a similar range
  of \lx--\Nh\ parameter space spanned by the \Swift-BAT and CDF-S
  samples.  See online manuscript for a colour version of this plot.}
\label{LXNH}
\end{figure}

The first step we take in defining the intrinsic infrared SEDs of
typical AGNs is to identify a sample of moderate luminosity (i.e.,
\lx$\sim10^{42}-10^{44}$ \ergs) AGNs that suffer minimal amounts of
host galaxy contamination at infrared wavelengths.  By doing this, we
only focus on those SEDs that require as little manipulation as
possible to extract the intrinsic AGN infrared SED.  In this section,
we describe how combine information derived from hard X-ray
observations and infrared spectra to produce such a sample.  We then
explain how we use \IRAS\ photometry to extrapolate the SEDs of a
subsample of these AGNs to FIR wavelengths.


The Burst Alert Telescope (hereafter, BAT) on-board the \Swift\
observatory is currently undertaking a survey of the sky at hard X-ray
energies (i.e., 14-195~\kev), where the effects from absorption (at
least to \Nh $< 10^{24}$~\cs) are negligible.  As a result, the second
data release of the \Swift-BAT survey, published in \cite{Tueller08},
provides a homogeneous sample of 154 local (i.e., $z<0.1$) X-ray AGNs.
A key motivation for using the AGNs selected from the \Swift-BAT
sample to define the intrinsic AGN infrared SED is that their X-ray
properties (i.e., 2-10~keV X-ray luminosities and absorbing columns,
hereafter, \lx\ and \Nh) cover largely the same ranges as those AGNs
detected in deep \Chandra\ surveys (e.g., CDF-N and CDF-S; see
\fig{LXNH}).  As such, this sample provides the ideal local analogue
to the high redshift AGNs detected in these fields as well as typical
local AGNs (i.e., \lx$\sim10^{42}-10^{44}$~\ergs\ and
\Nh$<10^{24}$~\cs; see \citealt{Winter09}).  Furthermore, in cases
where infrared photometry is available, the \Swift-BAT AGNs cover the
same range of infrared luminosities (integrated over 8-1000~\micron;
hereafter \lir; \lir$=6\times10^{9}-7\times10^{11}$~\lsun) as those
star-forming galaxies that make up the majority of the local galaxy
population detected by \IRAS\ (i.e., \lir$\sim10^{10}-10^{12}$~\lsun)
and $z<2$ galaxies detected in deep {\it Spitzer}/{\it Herschel}
fields (e.g. GOODS; P.I.s: M. Dickinson, D. Elbaz).

\begin{figure*}
\includegraphics[width=160mm]{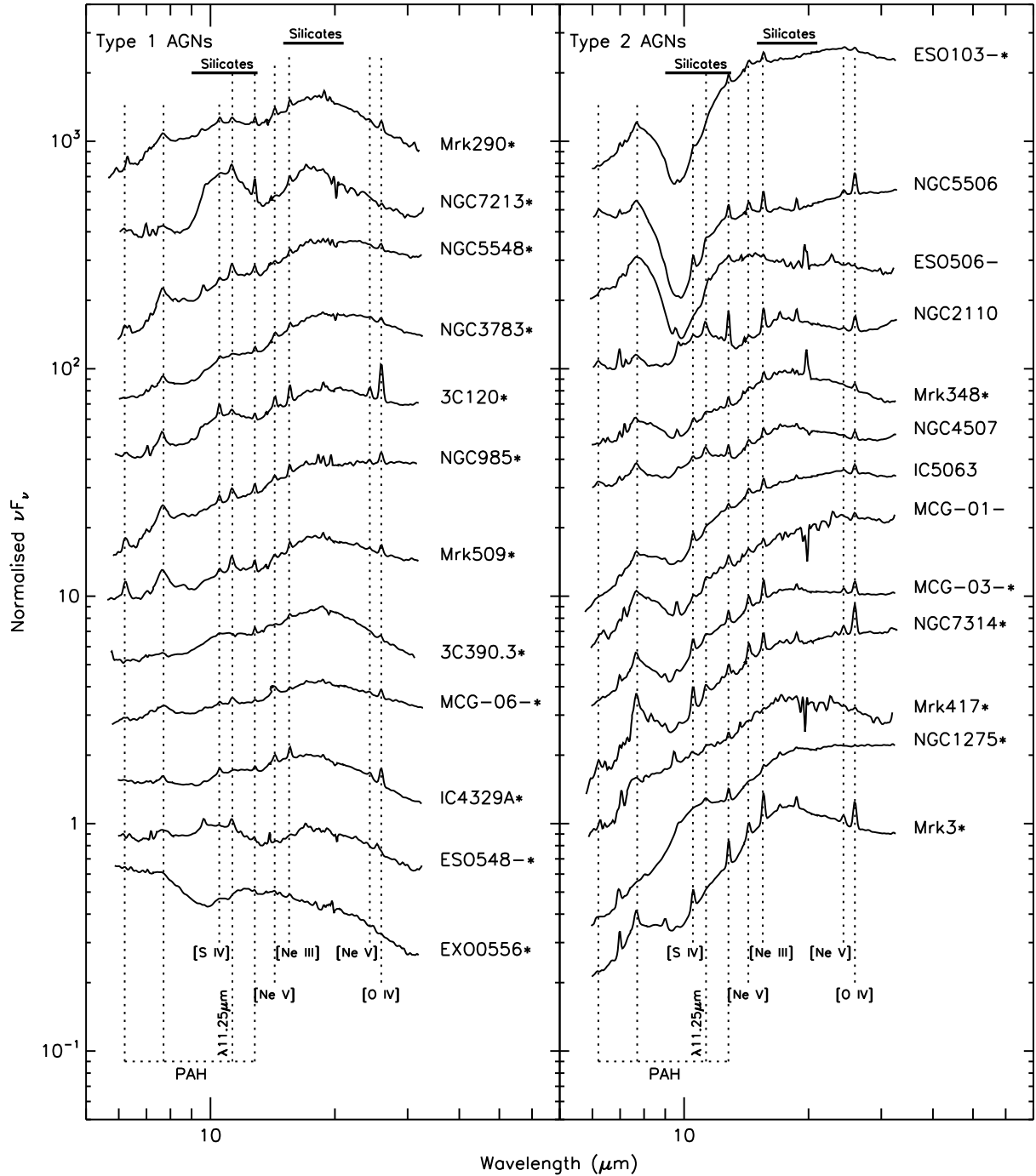}
\caption{The 25 AGNs from the \Swift-BAT sample that are strongly AGN
  dominated at mid-infrared wavelengths, separated in terms of their
  optical classification (here, Types 1.2 and 1.5 are classed as Type
  1 AGN, Types 1.8 and 1.9 are classed as Type 2 AGNs).  We have
  indicated the positions of the most common, prominent emission
  lines, PAH features and silicate emission/absorption features.  The
  criterion we use to determine whether the mid-infrared SED is
  AGN-dominated is that the equivalent width of the 11.25~\micron\ PAH
  feature is less than 0.03~\micron, which is equivalent to $<10$ per
  cent host-galaxy contribution at 19~\micron\ (e.g. \protect
  \citealt{Tommasin10}).  Within this subsample we see a wide range of
  mid-infrared SED shapes.  However, in the majority of cases (i.e.,
  at least 20/25; asterisked) the underlying AGN continuum can be
  described as an absorbed broken power law with a higher (i.e., more
  positive) spectral index longward of a break at, roughly,
  19~\micron.  In general, Type 2 AGNs have steeper SEDs (i.e., high
  spectral indices) at $\lambda\lesssim19~\micron$ although, in this
  respect, there is considerable overlap between the two classes.
  Type 2 AGNs are more likely to show evidence of strong silicate
  absorption at $\sim$10~\micron, while a larger proportion of Type 1
  AGNs have silicate in emission at $\sim$10~\micron\ and
  $\sim$18~\micron.  The galaxy names have been truncated to enable
  them to fit on the plot.  See table \ref{AGN_Dom_Table} for the full
  galaxy names.}
\label{AGNDominated}
\end{figure*}

In order to identify a sample of galaxies whose infrared output is
dominated by a moderate luminosity AGN, we cross-match the \Swift-BAT
sample with the archive of low resolution spectra obtained by the
infrared spectrograph (hereafter, IRS) on-board \Spitzer.  Of the 104
\Swift-BAT AGNs with measured \lx\ (see \citealt{Winter09}), 36 have
publicly available archival low resolution IRS spectra covering
(observed frame) $\sim6-35$~\micron. These spectra were reduced and
extracted following the procedures outlined in \cite{Mullaney10},
\cite{Goulding10} and Goulding (2010; PhD thesis).  Following the
diagnostics presented in \cite{Goulding09} and \cite{Tommasin10}, we
quantify the contribution of the host galaxy to these MIR spectra
using the equivalent widths of the 11.25~\micron\ PAH feature
(hereafter, \ewpah).  We decide to use the 11.25~\micron\ feature to
measure the host galaxy contribution rather than the typically
stronger one at 7.7~\micron\ as this latter feature is often blended
with another PAH emission line at 8.6~\micron.  We measure \ewpah\
using the spectral fitting code {\sc pahfit} (\citealt{Smith07}) and
assume that strongly AGN dominated systems have \ewpah $<$
0.03~\micron, corresponding to a $<10$ per cent host-galaxy
contribution at 19~\micron\ (see \citealt{Tommasin10}).\footnote{{\sc
    pahfit} is available from
  http://tir.astro.utoledo.edu/jdsmith/research/pahfit.php} Of the 36
X-ray AGNs that have archival low resolution IRS spectra, we identify
25 whose \ewpah\ satisfy this criterion (see \tab{AGN_Dom_Table} for a
list of these 25 AGN-dominated galaxies; also \fig{AGNDominated}).
Approximately half of these AGNs (i.e., 12/25) are optically
classified as either Type 1, 1.2 or 1.5 while the rest (i.e., 13/25)
are classified at either Type 1.8, 1.9 or 2 (optical classifications
taken from \citealt{Tueller08}; see \tab{AGN_Dom_Table} for these
classifications and other information regarding these 25 AGN dominated
galaxies).  We visually inspect the infrared spectra of each of these
25 AGNs to confirm that they are, indeed, dominated by a featureless
AGN continuum.

To provide constraints on the intrinsic AGN SED at FIR wavelengths, we
extrapolate beyond the low resolution IRS spectra using 60~\micron\
and 100~\micron\ photometry available from the \IRAS\ archives
(accessed through the NASA extragalactic database; NED).  To ensure
reliable extrapolation to FIR wavelengths, we only consider photometry
greater than 0.2~Jy and 1.0~Jy at 60~\micron\ and 100~\micron,
respectively (the approximate limiting sensitivity of the \IRAS\ faint
source catalogue at these wavelengths).  Of the 25 AGN-dominated
galaxies identified above, 20 satisfy these criteria.  However, we
note that aperture effects must be taken into consideration when
collating photometric and spectral data obtained by the \IRAS\ and
\Spitzer\ telescopes.  The apertures used to measure the \IRAS\
photometry reported in the point source and faint source catalogues
were between 2\arcmin\ and 9\arcmin\ in diameter, compared to the
3.6\arcsec\ and 10.5\arcsec\ (shortest dimension) apertures of the
short-low and long-low slits of the \Spitzer-IRS instrument,
respectively.  Therefore, there is considerable scope for the \IRAS\
photometry to contain a great deal more host galaxy flux than the
small aperture IRS spectra.  To mitigate such aperture effects, we
only extrapolate to the 60~\micron\ and 100~\micron\ photometry values
when the flux density in the IRS spectrum, integrated over the 12 and
25~\micron\ \IRAS\ band passes, matches (to within 30 per cent,
including errors) the \IRAS\ photometry at these wavelengths.  Our
choice of 30 per cent accuracy between the \IRAS\ and IRS fluxes is
based on the approximate systematic uncertainty of the \IRAS\ point
source and faint source catalogues when compared to the Revised Bright
Galaxy Sample (\citealt{Sanders03}).  We identify 11 AGN-dominated
galaxies that satisfy these criteria (indicated by a ``Y'' in column
16 of \tab{AGN_Dom_Table}), which we use to define the intrinsic AGN
SED.  The 6-100~\mum\ SEDs of these AGN-dominated galaxies, including
their 60~\mum\ and 100~\micron\ photometries, are presented in
appendix \ref{SEDFits}.

\pagestyle{empty}
\begin{landscape}
\begin{table}
  \begin{center}{
      \caption{Properties of the 25 AGNs in the \Swift-BAT sample that
        are strongly AGN dominated at 6-35~\micron.}
      \input{table1.tex}
\label{AGN_Dom_Table}   
}\end{center} {\sc Notes}: (1) Common name, (2) Optical Class from
\protect \cite{Tueller08} (3) and (4) Galaxy co-ordinates (J2000) from
\protect \cite{Tueller08}, (5) Distance in Mpc, $^1$denotes that a
redshift-independent distance measure taken from {\it NED} is used,
otherwise calculated from the redshift reported in \protect
\cite{Tueller08} (6) to (9) \IRAS\ photometry measurements, taken from
{\it NED} (10) and (11) 12~\micron\ and 25~\micron\ flux densities
derived from the low resolution IRS spectra, (12) Monochromatic
luminosity at 60~\micron\ in units of log(\lsun) (for comparison with
\protect \citealt{Netzer07}).  We use
\lsun$=3.83\times10^{33}$~\ergs. (13) Total infrared luminosity in
units of log(\lsun), calculated using the equations in Table 1 of
\protect \cite{Sanders96}, (14) Rest-frame 2-10~\kev\ luminosity,
corrected for absorption (taken from \protect \citealt{Winter09}) (15)
Equivalent width of the the 11.25~\micron\ PAH (i.e., \ewpah) feature
in \micron.  (16) Selected for extrapolation to $100$~\micron\ based
on whether the \IRAS\ photometry at 12 and 25~\micron\ matches that
derived from the IRS spectra. \end{table}
\end{landscape}

\begin{figure*}
\includegraphics[width=170mm]{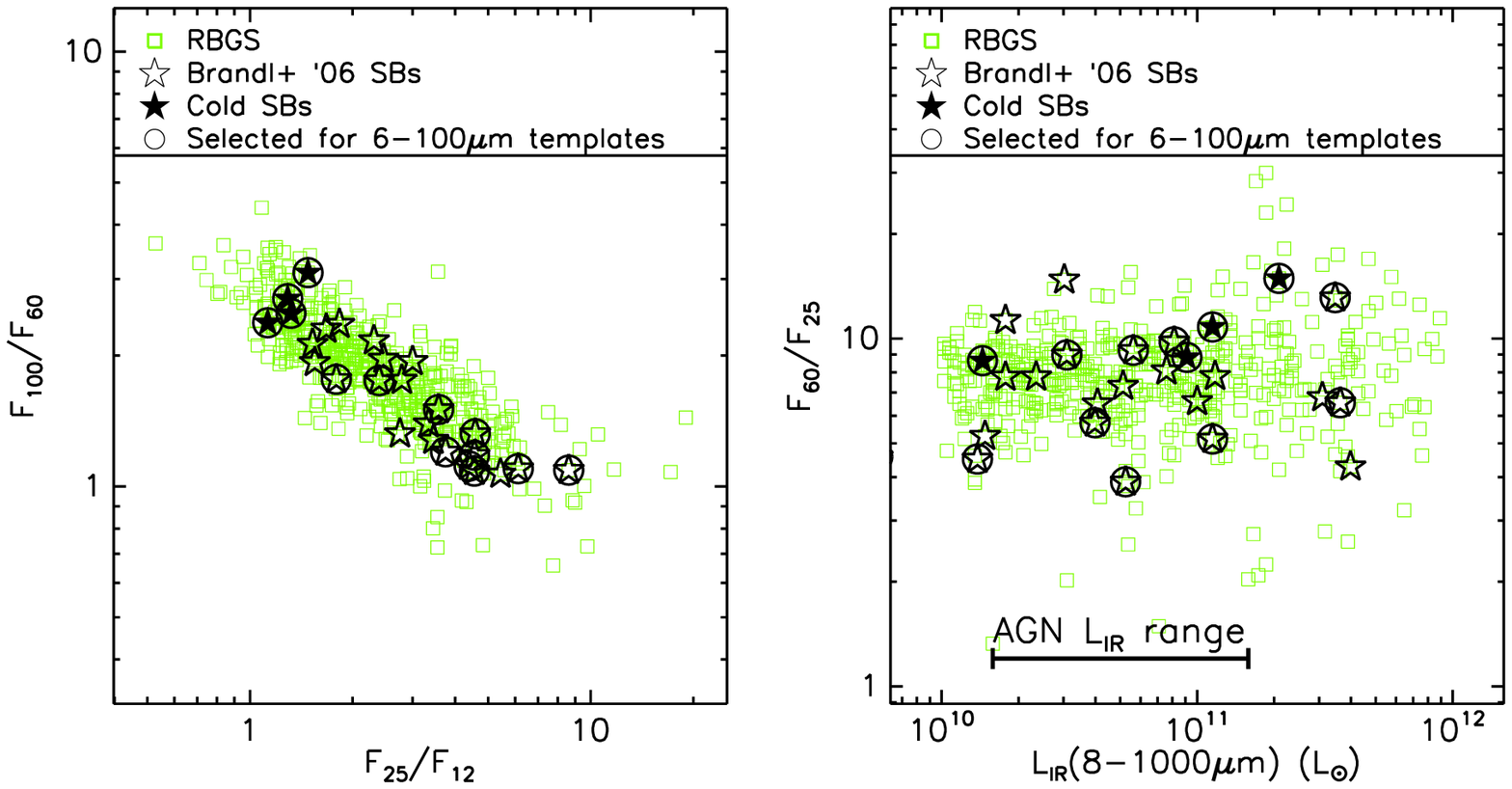}
\caption{\IRAS\ colour-colour (left panel) and colour-luminosity
  (right panel) plots of all sources in the {\it Revised Bright Galaxy
    Survey} (\protect \citealt{Sanders03}) with infrared luminosities
  covering the roughly the same range as the \Swift-BAT sample from
  which our AGN-dominated galaxies were selected (i.e.,
  \lir$=10^{10}-10^{12}$~\ergs).  We use a subsample of 10 starburst
  SEDs from \protect \cite{Brandl06} plus four other ``cold'' galaxies
  to define our host galaxy templates (circled stars).  These extra
  four quiescent-galaxy SEDs are needed to sample the full range of
  host-galaxy SEDs covered by \lir$=10^{10}-10^{12}$~\ergs galaxies.
  The lack of ``cold'' galaxies in the \protect \cite{Brandl06} sample
  is shown most clearly in the colour-colour plot (left hand panel).
  The horizontal black line in the right hand panel indicates the
  range of \lir\ of the AGNs selected for decomposition (see
  \S\ref{Templates:AGN:FIR}). The objects with
  $[$60~\micron/25~\micron$]\lesssim4$ are likely AGN dominated and
  were therefore not considered as suitable host-galaxy templates.}
\label{RBGS}
\end{figure*}

\section{Defining a set of host-galaxy templates}
\label{Templates:SB}

The sample of AGNs identified in the previous section show minimal
amounts of host-galaxy contamination at MIR wavelengths.  However,
because a typical host-galaxy SED rises strongly at longer infrared
wavelengths, an SED that is strongly AGN-dominated at MIR wavelengths
may not necessarily be AGN-dominated at FIR wavelengths.  Indeed,
\cite{Netzer07} showed that at 60~\mum\ the average quasar SED is 80
to 90 per cent host-galaxy dominated, despite having only weak PAH
features at MIR wavelengths. In this section we describe the
construction of a set of host-galaxy templates that represent the full
diversity of host-galaxy SEDs expected for typical AGNs (i.e., the
full range of IRAS colour-colour space for
\lir$=10^{10}-10^{12}$~\lsun\ galaxies). These templates are used to
account for the host-galaxy contribution to our AGN-dominated sample
in \S\ref{Templates}.

In \fig{RBGS} (left panel) we plot the $[$100~\micron/60~\micron$]$
vs. $[$25~\micron/12~\micron$]$ colours of all
\lir$=10^{10}-10^{12}$~\lsun\ galaxies in the {\it Revised Bright
  Galaxy Survey}.  From this figure it is clear that the majority of
the colour-colour space spanned by these galaxies is well sampled by
the \cite{Brandl06} sample of starburst galaxies.  However, we note
that there is a region of this colour-colour space that is not sampled
by these starbursts.  Galaxies in this un-sampled region of the
colour-colour space have SEDs that rise steeply between 60~\micron\
and 100~\micron, but are relatively flat between 12~\micron\ and
25~\micron.  As such, they likely represent a population of cold
galaxies.  To construct our host-galaxy templates we use a selection
of \cite{Brandl06} starbursts and a selection of galaxies that lie in
this ``cold'' region of the \IRAS\ colour-colour space.  This ensures
that we cover the full range of SEDs displayed by
\lir$=10^{10}-10^{12}$~\lsun\ galaxies.  We note that the both the
colour-colour and $[$60~\micron/25~\micron$]$-\lir\ parameter spaces
are particularly well-sampled in the infrared luminosity range spanned
by the AGNs selected for decomposition (i.e.,
log(\lir/\lsun)$=10.2-11.2$~\lsun; \fig{RBGS}, right panel; see
\S\ref{Templates:AGN:FIR}).

We extracted and reduced the low resolution IRS spectra of all the
\cite{Brandl06} starbursts and four additional ``cold'' galaxies
(namely, NGC~1667, NGC~5734, NGC~6286 and NGC~7590), following the
same procedures as those referred to in \S\ref{Sample}.  All of the
\cite{Brandl06} starbursts were observed in IRS staring mode, while
the four ``cold'' galaxies were observed in mapping mode.  We note
that these differences in the observing modes will have little or no
effect on our host-galaxy templates as our selection process minimises
aperture effects (see later).  We extrapolated these MIR SEDs to
100~\micron\ using \IRAS\ photometry as described in
\S\ref{Sample}. To mitigate aperture effects we only use those SEDs in
which the IRS spectra agree with the 12~\micron\ and 25~\micron\
\IRAS\ photometry to within 30 per cent. Out of the 16 \cite{Brandl06}
starbursts identified as either pure starbursts or starburst$+$LINER,
we identify 10 that satisfy these criteria: Mrk~52, Mrk~520, NGC~660,
NGC~1222, NGC~2623, NGC~3256, NGC~4194, NGC~4818, NGC~7252,
NGC~7714. All four of the ``cold'' galaxies also satisfy these
criteria.  To produce our host-galaxy templates we group these SEDs in
terms of their overall shape and the relative strength of their PAH
features (see \fig{SBs}). The full range of host-galaxy SED shapes is
well characterised by five groups of SEDs, referred to as ``SB1''
through ``SB5'', with all four ``cold'' SEDs contained within a single
group (i.e., ``SB1''; see \tab{SB_Table} and \fig{SBs}). Five
host-galaxy templates are obtained by normalising each SED within each
group at 90~\micron\ and calculating their mean average.  Since the
SEDs within each group are so similar, the wavelength at which they
are normalised has little effect on these host-galaxy templates.
Beyond 100~\micron\ we extrapolate the average SEDs as a modified
black body (i.e., $F_\nu=F_\nu^{\rm BB}\nu^{\beta}$, where $F_\nu^{\rm
  BB}$ is the blackbody specific flux and $\nu$ is photon frequency;
we adopt $\beta=1.5$).  We verify that this is a reasonable assumption
using AKARI 140~\micron\ and 160~\micron\ photometry data available
for 13 of the 14 host-galaxy SEDs in our sample.\footnote{AKARI data
  were retrieved from the NASA/IPAC Infrared Science Archive at
  http://irsa.ipac.caltech.edu/.} These templates are used to remove
any host-galaxy contribution to the infrared SED of the AGN-dominated
galaxy sample.  We publish the host-galaxy templates in columns 5-9 of
\tab{SED_Table} (which is available in its entirety online at
http://sites.google.com/site/decompir).

\section{Characterising the intrinsic infrared AGN SED}
\label{Templates}

First we use the IRS spectra of our AGN sample (defined in \S2.1) to
explore the range of AGN-dominated infrared SEDs at 6-35~\mum.  We
then extract the intrinsic 6-100~\mum\ SEDs of a subsample of these
AGNs by carefully decomposing the observed SEDs into their host-galaxy
(defined in \S\ref{Templates:SB}) and AGN components.

\subsection{AGN dominated SEDs at 6~\micron\ to 35~\micron}
%
\label{Templates:AGN:MIR}

The IRS spectra presented in \fig{AGNDominated} provide a clear
picture of the range of MIR SEDs produced by strongly AGN-dominated
systems.  As explained in \S\ref{Sample}, the weak PAH features in
this sample suggest that at least 90 per cent of the continuum
emission at 19~\micron\ is produced by the AGN.  Considering that the
typical host-galaxy continuum falls toward shorter infrared
wavelengths (see \fig{SBs}), it is reasonable to assume that the AGN
dominates the continuum emission at $\sim$6~\micron\ -- 20~\micron. At
these wavelengths, all 25 AGN-dominated SEDs show clear evidence of a
continuous, underlying power-law continuum that is thought to be
produced by multiple dust components spanning a range of temperatures
(e.g., \citealt{Buchanan06}).
  
The spectral indices of the underlying AGN power-law continua span the
range $0.7\leq\alpha_1\leq2.7$ (mean: $\alpha_1$=1.6), with a tendency
for Type 2 AGNs to have higher (more positive) spectral indices
compared to Type 1 AGNs. \footnote{Here, we use $F^{\rm
    AGN}_\nu\propto\lambda^{\alpha}$.  To convert to $\nu F^{\rm
    AGN}_\nu $ use $\beta = \alpha - 1$, where $\nu F^{\rm
    AGN}_\nu\propto\lambda^{\beta}$} However, there is considerable
overlap between the types (i.e., Type 1s: $0.7\leq\alpha_1\leq1.7$;
Type 2s: $0.8\leq\alpha_1\leq2.7$), and the difference in the mean
SEDs is not statistically significant. This large range of spectral
indices is consistent with those found in previous studies
(e.g. \citealt{Buchanan06,Wu09}). In four cases (i.e., one Type 1:
EXO055620-3820.2 and three type 2s: ESO506-G027, NGC~5506,
ESO103-035), we see evidence of strong silicate absorption at
9.7~\micron.  In a further eight cases (i.e., five type 1s:
ESO548-G081, 3C120, NGC~3783, Mrk~290, NGC~7213 and three type 2s:
NGC~1275, NGC~2110, NGC~4507) there is strong evidence of silicate
emission at $\sim10$~\micron\ which is always accompanied by another
silicate emission feature at $\sim18$~\micron\ (see \citealt{Sturm05}
for a dedicated study of the silicate emission features seen in the
MIR spectra of AGNs).

In at least 20 of the 25 AGN-dominated MIR spectra there is a definite
break in the power-law continuum at $15-20$~\micron; mean break
position of $\sim19$~\micron.  In previous studies this has been
attributed to a dominating warm (i.e., $\sim170$~K) dust component
that is heated by the AGN (e.g., \citealt{Weedman05, Buchanan06}). In
all cases where we see a break in the continuum power law, the AGN SED
longward of $\lambda_{\rm Brk}$ has a lower spectral index (i.e.,
power-law index, $0\leq\alpha_2\leq1.5$, mean$=0.7$; again, consistent
with \citealt{Buchanan06} and \citealt{Wu09}) than at shorter
wavelengths.  We again find that there is significant overlap in the
range of spectral indices at $\lambda>19$~\micron\ between Type 1 and
Type 2 AGNs. We note that the power-law indices shortward and longward
of $\lambda_{\rm Brk}$ are uncorrelated.  The number of AGNs in our
sample showing a break around 19~\micron\ could be as high as 24,
since it can be disguised by the presence of the silicate emission
feature at 18~\micron.  The only AGN in our sample where the
6--35~\mum\ SED can be unambiguously described as a single power law
is NGC5506; the spectral index of NGC5506 over 6--35~\mum\ is
$\alpha=1.2\pm0.2$, which is consistent with the ranges of both
$\alpha_1$ and $\alpha_2$.

Because the typical host-galaxy continuum emission increases strongly
toward longer wavelengths, it is difficult to ascertain whether the
power-law emission at longer wavelengths (i.e.,
$\lambda\gtrsim25$~\mum) arises from the AGN alone, or if the host
galaxy also makes a significant contribution.  Indeed, in four cases
(i.e., NGC7213, NGC4507, NGC2110, 3C120), the SED longward of
$\sim25$~\micron\ shows evidence of a turn-up in $\nu F_\nu$ which is
consistent with being due to an underlying host-galaxy component.  The
host-galaxy component must be accounted for at longer wavelengths in
order to define the intrinsic AGN infrared SED, which we address in
the next subsection.

\subsection{Intrinsic AGN infrared SED at 6~\micron\ to
  100~\micron}

\label{Templates:AGN:FIR}

We now extend the intrinsic AGN infrared SEDs to longer wavelengths
(i.e., $\lambda\gtrsim25$~\mum) using IRAS 60~\mum\ and 100~\mum\
photometry for guidance. Here, we only consider those 11 AGN-dominated
SEDs that we have shown suffer from minimal aperture effects (see
\S\ref{Sample}).  The approach we take is to decompose the observed
SEDs into their host-galaxy and intrinsic AGN components through
simultaneously fitting the IRS spectra and IRAS photometry.  An
alternative approach would be to simply subtract a normalised
host-galaxy component from the observed SEDs to leave the emission
intrinsic to the AGN.  However, calculating the appropriate
normalisation for the host-galaxy component is a non-trivial matter
that is complicated by the fact that the host-galaxy SED may be
modified by dust-absorption (i.e., the observed SED is {\it not} a
simple linear sum of a host-galaxy and an AGN component; separate
absorption terms must also be considered).  Later, and only for
illustrative purposes, we subtract the host-galaxy SEDs normalised via
our SED fitting routine from the sample of 11 observed AGN SEDs (see
\fig{IntrinsicSEDs}).

One downside of the approach that we adopt is that it requires us to
make {\it a priori} assumptions about the form of the intrinsic AGN
infrared SED used in our fits.  Our analyses of the AGN-dominated MIR
spectra in the previous subsection provides us with a well defined set
of parameters that describe the variety of intrinsic AGN SEDs at 6 to
$\sim$25~\micron\ (i.e., $0.7\leq\alpha_1\leq2.7$,
$0\leq\alpha_2\leq1.5$ and
15~\micron$\leq\lambda_{Break}\leq$20~\micron).  At longer
wavelengths, where there are fewer observational constraints, we can
only estimate the form of the intrinsic SED based on reasonable
physical assumptions then test whether models incorporating these SEDs
reproduce observations.  With this in mind, we assume that the
intrinsic SED falls as a modified black body beyond a given
wavelength, $\lambda_{\rm BB}$.  In our fits, $\lambda_{\rm BB}$ is
allowed to take any value between 20~\micron\ and 100~\micron.  Our
choice of a modified black body spectrum is loosely based on the
results of radiative transfer models of the dust surrounding AGNs,
although we note that the precise form of this fall-off does not have
a significant impact on the fits.  When needed, we add silicate
emission features at 10~\micron\ and 18~\micron\ to the intrinsic AGN
component.  These are modelled with broad (i.e., FWHM$\sim$3~\micron)
gaussians. In our models, we account for any absorption to both the
AGN and host-galaxy components using a \cite{Draine03} extinction
curve.

\begin{table}
  \caption{Galaxy SEDs that used to construct our host galaxy templates.}
  \begin{center}
  \input{table2.tex}
  \end{center}
\label{SB_Table} 
\medskip
{\sc Notes}: (1) Host-galaxy template, (2) Galaxy name, (3) Distance in Mpc taken from \protect \cite{Sanders03}, $^a$calculated using the redshift reported in NED using the same cosmology as \protect \cite{Sanders03}, $^b$redshift-independent distance reported in NED 
(4) 8--1000~\micron\ infrared luminosity, taken from \cite{Sanders03}, $^{c}$calculated using the distance in column 3 and the prescription outlines in table 1 of \protect \cite{Sanders96}. 
\end{table}

\begin{figure}
\includegraphics[width=85mm, height=140mm]{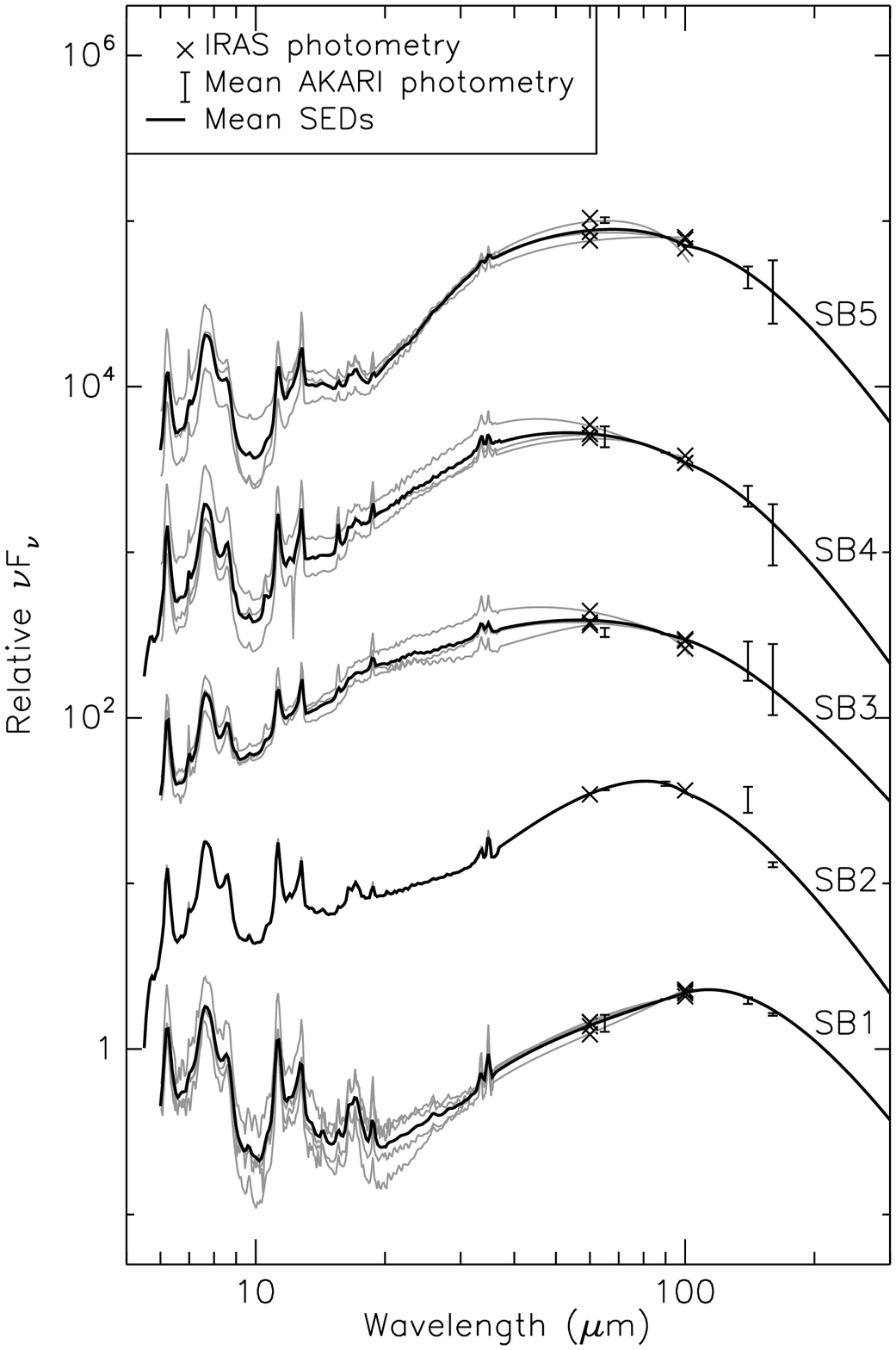}
\caption{The five average host-galaxy templates derived from the
  \protect \cite{Brandl06} starburst sample and the four ``cold''
  galaxies selected from the {\it Revised Bright Galaxy Sample}
  (\protect \citealt{Sanders03}; see \fig{RBGS}).  Each template is
  the mean-average of a group of SEDs that have similar overall shape
  and relative PAH strengths (see \S\ref{Templates:SB} for a
  description of how we construct these host-galaxy templates).  Each
  SED is made up of a low-resolution IRS spectrum that we extrapolate
  to far-infrared wavelengths using \IRAS\ photometry at 60~\micron\
  and 100~\micron. We also show the mean flux at 65~\micron,
  140~\micron\ and 160~\micron\ measured from AKARI data (when
  available) as a check that the modified black-body extrapolation
  beyond 100~\micron\ is a reasonable approximation of the real SEDs.
  We use these average templates to model any host galaxy components
  in our sample of AGN-dominated galaxies to derive the intrinsic AGN
  infrared SED.}
\label{SBs}
\end{figure}

\begin{figure*}
\includegraphics[width=180mm]{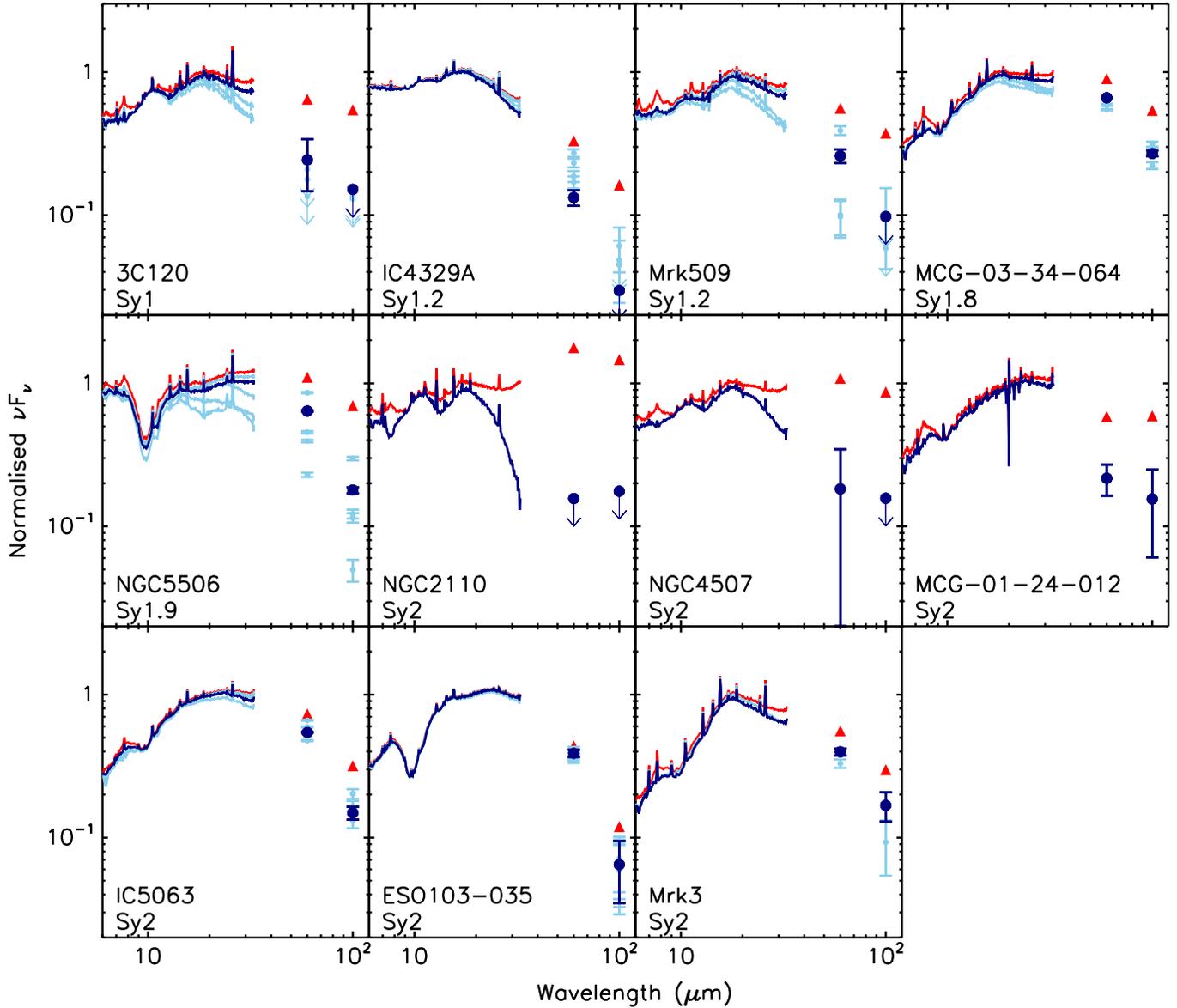}
\caption{The intrinsic mid-infrared spectra (light blue lines) and
  far-infrared photometry (light blue points) of the 11 AGN-dominated
  sources in our sample after subtracting suitable host-galaxy
  components.  The suitability of the host-galaxy components is
  assessed from the SED fits described in \S\ref{Templates:AGN:FIR}
  (see also appendix \ref{SEDFits}).  We highlight the SED produced by
  subtracting the most suitable host-galaxy component (i.e., the
  ``best-of-the-best''; see \S\ref{Templates:AGN:FIR}; dark blue
  line/points).  The original observed data is shown in red.  The
  intrinsic AGN infrared SEDs shown here cover a broad range of
  shapes, although all fall rapidly at wavelengths longwards of
  15-60~\mum.  Despite this fall-off at long wavelengths, there are at
  least four observed AGNs SEDs (namely, Mrk~3, MCG-03-34-064,
  ESO-103-035 and IC5063) that are AGN-dominated even at 60~\mum,
  irrespective of the choice of subtracted host-galaxy template.}
\label{IntrinsicSEDs}
\end{figure*}

We proceed by simultaneously fitting the IRS spectra and the IRAS
photometry of the 11 AGNs in our sample with a combination of our
host-galaxy templates (see \S\ref{Templates:SB}) and the estimated
intrinsic AGN SED described above.  Any strong emission lines
(excluding PAH features) are masked to ensure that they do not
adversely effect the fit.  Each of the 11 infrared SEDs are fit five
times, each time using a different host-galaxy template (defined in
\S\ref{Templates:SB}; see also \fig{SBs}).  We use
$\chi^2$-minimisation to obtain the best fitting parameters for each
choice of host-galaxy template.  Plots showing the fits to the data
(including the host-galaxy and intrinsic AGN components) and the
resulting residuals are presented in appendix \ref{SEDFits}.  

Because the observed SEDs consist, in part, of very high
signal-to-noise IRS spectra, none of the fits to the data are formally
good (i.e., their reduced $\chi^2>>1$).  Therefore, we cannot use
$\chi^2$ statistics to unambiguously determine whether any of the five
different AGN--host-galaxy template solutions provide a good
characterisation of the data.  Instead, we use the following criteria
to determine which of the five models provide suitable fits to the
observed data: does it (a) pass within 2$\sigma$ of the 60~\mum\ and
100~\mum\ photometry and (b) reproduce the general shape of the IRS
spectrum.  For each of the 11 SEDs in our sample, we can identify at
least one of the five model fits that satisfy both these criteria.
When both criteria are met we select, by eye, those model fits that
produce the smallest residuals at the wavelengths of the strongest PAH
features, i.e., 7.7~\mum, 11.25~\mum\ and 12.8~\mum.  We use this
additional criterion to ensure that any contribution from the
host-galaxy well accounted-for by the fitted host-galaxy
component. Where there is no significant difference in the PAH
residuals of two or more such fits we identify each of them as being
suitable (in appendix \ref{SEDFits} we have highlighted all selected
fits).  In only two cases, namely NGC~5506 and ESO103-035, do we see
large residuals at the PAH wavelengths for all five fits.  In both
these cases, we assume that all of the fits are as good as each other
and select all five for further consideration.  In all selected cases
(including NGC~5506 and ESO103-035), the model fits lie within 10 per
cent of the measured flux density for at least 85 per cent of the IRS
data points, demonstrating the ability of our SED fitting approach to
reproduce the continuum shape and PAH strengths of the AGNs in our
sample. As a final note to the fitting procedure we would like to
  point out that, while we are confident that selecting the fits that
  closely reproduce the PAH features is the most appropriate procedure
  to take, our main results do not change if, instead, we use the
  results derived from using all our host-galaxy templates (which
  cover the full range of SED shapes of local,
  \lir$=10^{10}-10^{12}$~\lsun\ galaxies; see \fig{RBGS}).  The
  strongest effect that taking this alternative approach has on our
  results is to increase the scatter of intrinsic AGN SEDs at FIR
  wavelengths by a factor of $\sim2-3$; as we shall see, this does not
  represent a large increase over the range of intrinsic AGN SED
  shapes produced by selecting only the suitable host-galaxy
  components identified using the selection criteria outlined above.

For illustrative purposes, we show in \fig{IntrinsicSEDs} the SEDs
obtained by subtracting the suitable host galaxy components (i.e.,
those extracted from the suitable model fits identified above and
highlighted in appendix \ref{SEDFits}) from the observed SEDs.  As we
have selected only those fits that reproduce the general shape of the
observed SED, the SEDs presented in \fig{IntrinsicSEDs} are equivalent
to the intrinsic AGN components produced during the fitting procedure
and shown in appendix \ref{SEDFits}.  For guidance, we have
highlighted the SED that is produced by the fit with the smallest
overall residuals between 6~\mum\ and 35~\mum\ (i.e., the
``best-of-the-best''), although we stress that all plotted intrinsic
SEDs are viable.

For all 11 galaxies in our sample, the choice of host-galaxy template
has almost no effect on the derived intrinsic SED shortward of
$\lambda_{\rm Brk}$.  On the other hand, in at least five cases the
SED at longer wavelengths is strongly dependent on the choice of
host-galaxy template, meaning the exact form of the intrinsic AGN SED
at these wavelengths is uncertain.  However, in all cases the
intrinsic SED falls rapidly (in $\nu F_\nu$) beyond 15-60~\mum,
irrespective of the choice of host-galaxy component.  Despite this
fall-off at long wavelengths, the total emission at 60~\mum\ is still
dominated by the AGN in at least three, possibly four, cases (namely,
MCG-03-34-064, ESO-103-035, IC5063 and, possibly, Mrk~3; there is one
unselected host-galaxy component that dominates at 60~\mum\ in this
last case), irrespective of the choice of host-galaxy SED (see
\fig{IntrinsicSEDs}).

\begin{figure}
\includegraphics[width=83mm]{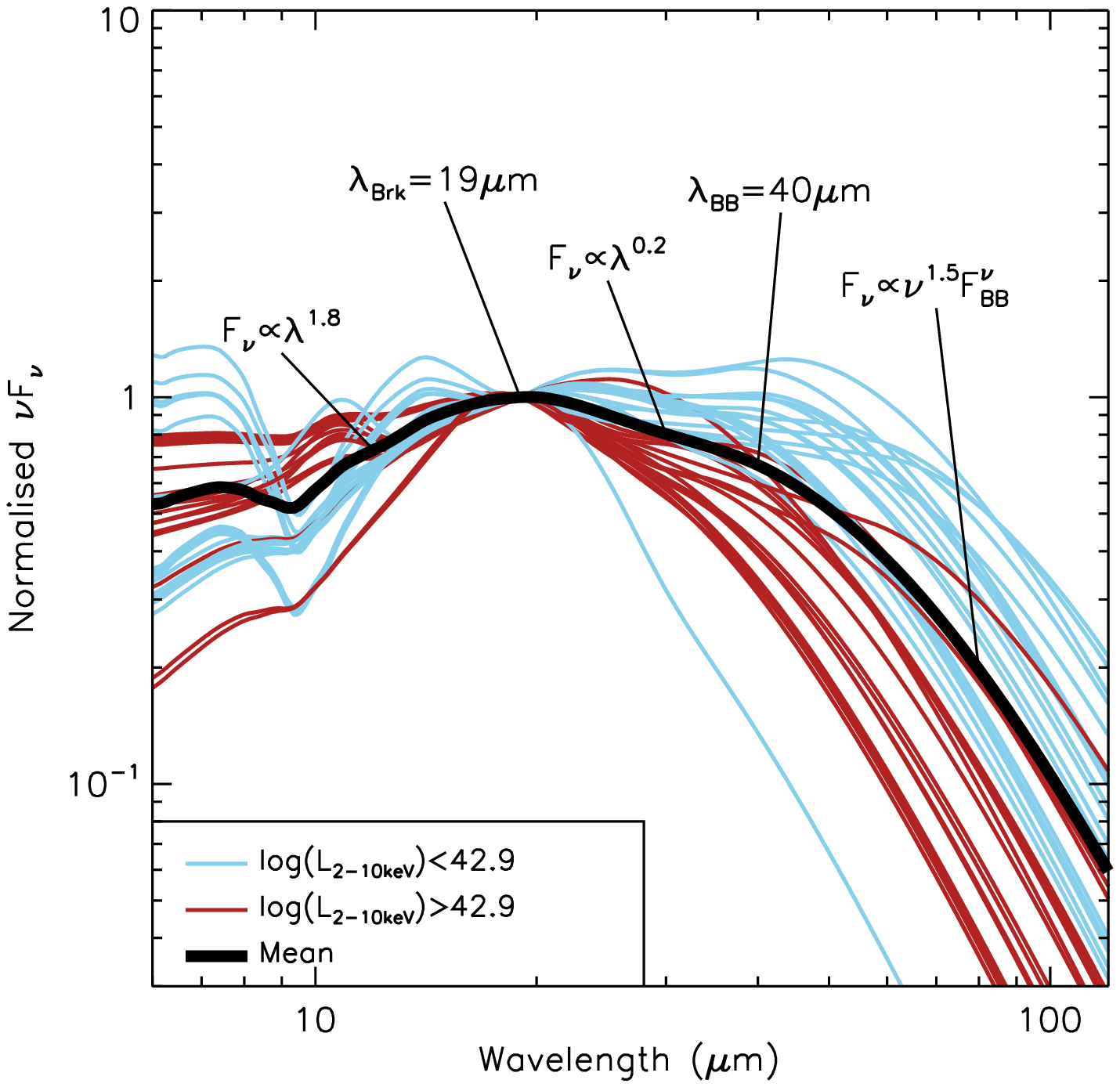}
\caption{The full range of possible intrinsic AGN infrared SEDs
  produced by our SED fitting procedure described in
  \S\ref{Templates:AGN:FIR}.  We include in this plot the intrinsic
  SEDs extracted from the fits deemed suitable using the criteria
  outlined in \S\ref{Templates:AGN:FIR} (i.e., those highlighted in
  appendix \ref{SEDFits}). Each intrinsic SED is normalised at
  19~\mum\ to demonstrate the range of SED shapes.  There is a clear
  systematic difference between the intrinsic SEDs of AGNs with X-ray
  luminosities above and below the median of our sample (i.e., ${\rm
    log}($\lx$)=$42.9).  Also included in this plot is the
  mean-average intrinsic SEDs of all 11 AGNs in our sample, labelled
  to illustrate the parameterisation outlined in Eqn. \ref{Eqn:AGN}.}
\label{Range}
\end{figure}

In \fig{Range} we plot the full range of possible intrinsic AGN SED
components extracted from our models.  Shown in this plot are the
intrinsic SEDs from all suitable fits identified using the criteria
outlined above.  A notable feature of this plot is the large range of
intrinsic SED shapes extracted from the fits to the 11 observed SEDs
in our sample.  For example, when normalised at 19~\mum\ the flux the
intrinsic AGN flux extracted from our model fits spans over an order
of magnitude at 60~\mum\ and 100~\mum.  While the range mid-infrared
(i.e., $\lambda\lesssim 25$~\mum) SEDs is well constrained by the
AGN-dominated IRS spectra, it is not clear how much of the scatter at
FIR wavelengths (i.e., $\lambda\gtrsim 25$~\mum) is due to differences
in the true intrinsic SED and how much is introduced by the different
host-galaxy templates we use.  Therefore, the range shown here will
cover a larger spread than the true intrinsic AGN SEDs at
$\lambda\gtrsim25$~\mum.  In \fig{Range} we have discriminated between
AGNs with X-ray luminosities above and below the median of the sample
(i.e., above and below ${\rm log}($\lx$)=$42.9).  We identify a clear
trend for more intrinsically luminous AGNs to have intrinsic infrared
SEDs that fall more rapidly at longer wavelengths, although stress
that there is some overlap between the SEDs of AGNs with X-ray
luminosities above and below the median of our sample (indeed, the
most rapidly falling SED is that of the lowest luminosity AGN in our
sample, NGC~2110).  We also find that, on average, the mid-infrared
SEDs of the more X-ray luminous AGNs in our sample have weaker
silicate absorption and stronger silicate emission, although this
could be related to the fact that all of the low X-ray luminosity AGNs
are classed as either Type 1.8, 1.9 or 2 AGNs which, as we have seen,
tend to have stronger silicate absorption features (see
\S\ref{Templates:AGN:MIR}).  The evidence for a link between the
shape of the intrinsic infrared SED and \lx\ is strengthened by
differences between the average intrinsic SEDs of the moderate
luminosity AGNs studied here and those of more luminous quasars (e.g.,
the average quasar SED presented in \citealt{Netzer07}; see
\S\ref{Comparison:QSO}).  There is weak evidence of a relation between
the shape of the SEDs shortward and longward of $\lambda_{\rm Brk}$,
with SEDs with lower values of $\alpha_1$ falling more rapidly at FIR
wavelengths.  However, we note that there are at least two objects in
our sample that deviate strongly from this trend (namely, Mrk~3 and
NGC~5506; see \fig{IntrinsicSEDs}).

\begin{figure*}
\includegraphics[width=170mm]{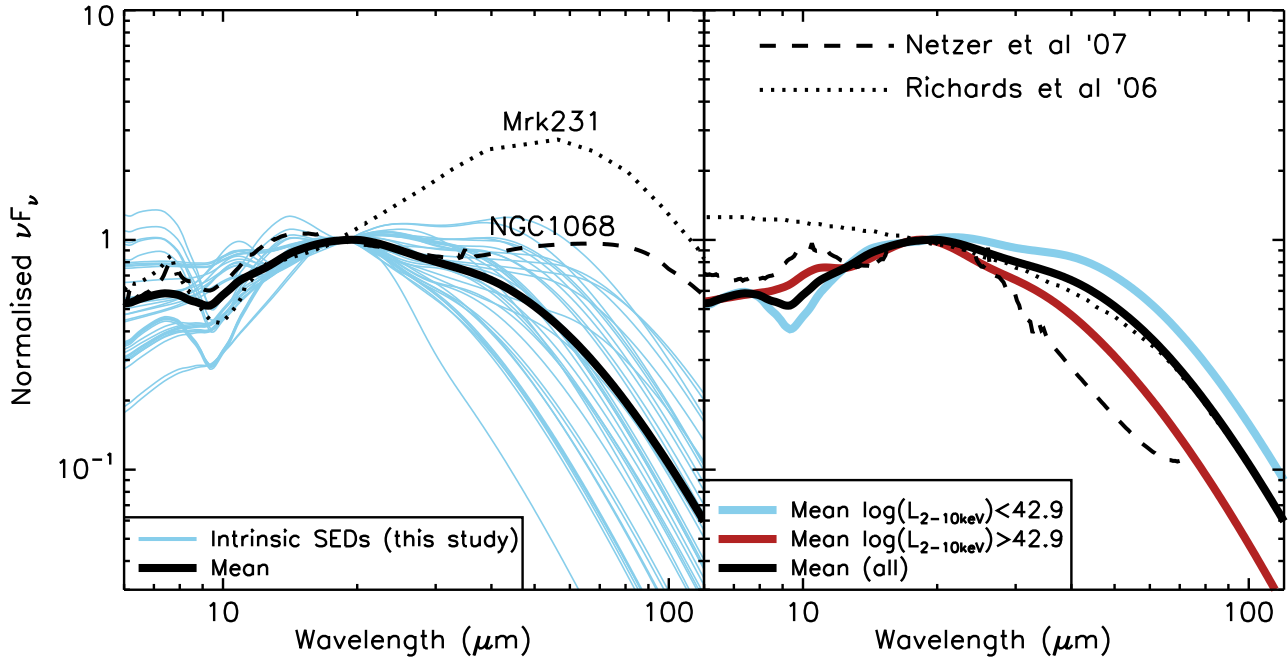}
\caption{{\it Left}: The mean-average and range of intrinsic AGN
  infrared SEDs from this study plotted against the observed infrared
  SEDs of Mrk~231 and NGC~1068 (i.e., two galaxies commonly used to
  represent the typical AGN infrared SED).  When all are normalised at
  19~\mum, the SEDs of NGC~1068 and Mrk~231 lie above the intrinsic
  AGN SEDs at FIR wavelengths, which we attribute to host-galaxy
  contamination in the two comparison SEDs (see
  \S\ref{Comparison:NGC1068}). {\it Right}: The mean average intrinsic
  SEDs of (a) all 11 AGNs in our sample, (b) log(\lx)$<$42.9 AGNs and
  (c) log(\lx)$>$42.9 AGNs, plotted against the average quasar SEDs of
  \protect \cite{Richards06} and the average intrinsic quasar infrared
  SED of \protect \cite{Netzer07}.  Again, all SEDs are normalised at
  19~\micron.  Note that the \protect \cite{Richards06} and \protect
  \cite{Netzer07} SEDs extend to $\sim95$~\micron\ and
  $\sim$65~\micron, respectively.  At this normalisation, the \protect
  \cite{Richards06} average quasar SED is well matched to the average
  intrinsic SED at $\lambda\gtrsim$30~\mum, although it lies above our
  average intrinsic SED at shorter wavelengths.  At
  $\lambda\lesssim$19~\mum\ the average intrinsic quasar SED of
  \protect \cite{Netzer07} lies slightly above all the average
  intrinsic SEDs from this study, but lies below them at
  $\lambda\gtrsim$19~\mum.  This plot clearly shows the trend reported
  in \S\ref{Templates:AGN:FIR} for more luminous AGNs to have bluer
  intrinsic infrared SEDs.}
\label{CompareObs}
\end{figure*}

We also plot in \fig{Range} the average intrinsic SEDs calculated by
taking the mean of the intrinsic AGN components extracted from the
fits.  Because of the considerable overlap in shape of the intrinsic
SED at MIR wavelengths between Type 1 and Type 2 AGNs (see
\S\ref{Templates:AGN:MIR} and \fig{AGNDominated}) and the small number
of intrinsic SEDs in our sample (i.e., 11) we do not discriminate
between AGN types when producing these averages.  Although not
included in this plot, we note that the average of the SEDs obtained
by subtracting the host-galaxy components from the observed data
(i.e., those shown in \fig{IntrinsicSEDs}) is consistent with the
average SED shown in \fig{Range}. This average SED can be expressed
as:
\begin{equation}
F_\nu \propto \left\{
\begin{array}{rcl}
  \lambda^{1.8} & \mbox{at} & 6~\umu{\rm m} < \lambda < 19~\umu{\rm m}\\
  \lambda^{0.2} & \mbox{at} & 19~\umu{\rm m} < \lambda < 40~\umu{\rm m} \\
  \nu^{1.5}F_\nu^{\rm BB} & \mbox{at} & \lambda > 40~\umu{\rm m} \\
\end{array}\right.
\label{Eqn:AGN}
\end{equation}
\noindent
and shows evidence of weak silicate absorption (equivalent to an
absorbing column of \Nh$\sim$5\e{21}\cs, or $\tau_{9.7}\sim$0.2, using
fig. 10 of \citealt{Draine03} to convert \Nh\ to $\tau_{9.7}$) but
little or no silicate emission. For clarity, we plot the average
mean-average SEDs of ${\rm log}($\lx$)>$42.9 and ${\rm
  log}($\lx$)<$42.9 AGNs separately in the right-hand panel of
\fig{CompareObs}.  As expected from the trend identified above, the
average intrinsic SED of the more luminous AGNs in our sample (i.e.,
${\rm log}($\lx$)>$42.9) falls more rapidly at longer wavelengths than
that of the lower luminosity AGNs (i.e., ${\rm log}($\lx$)<$42.9).
For example, when normalised at 19~\mum\ the lower luminosity AGNs
emit, on average, 2-3 times more flux at 60~\mum\ and 100~\mum\ than
the higher luminosity AGNs in our sample.  The parameters describing
the average SEDs of the high and low luminosity AGNs in our sample are
largely the same as shown in Eqn. \ref{Eqn:AGN}, although the spectral
indices at $\lambda>19$~\mum\ are somewhat different, i.e.,
$\alpha_2$=0.0 and 0.4, respectively.  This relative difference
between the FIR SEDs of high and low luminosity AGNs in our sample
could be the result of higher luminosity AGNs being capable of heating
a larger fraction of their surrounding dust to higher temperatures.  A
consequence of this increased heating would be relatively stronger
emission at MIR wavelengths; hence the apparent flux deficit at FIR
wavelengths when normalised at 19~\mum\ (see also
\S\ref{Comparison:QSO}).  All three average intrinsic AGN SEDs
described here are published in columns 2-4 of \tab{SED_Table} (which
is available in its entirety online at
http://sites.google.com/site/decompir).

Finally, we explore whether using a more straightforward approach to
normalise the host galaxy component has any effect on the extracted
intrinsic AGN SEDs.  To extract the average intrinsic SED of more
luminous, quasar AGNs, \cite{Netzer07} normalised the host galaxy
component such that it constitutes the majority of the observed flux
at 60~\mum\ and 100~\mum.  If we take a similar approach and assume
that 90\% of the flux at 100~\mum\ is emitted by the host-galaxy, we
obtain largely the same overall intrinsic SED shapes as shown in
\fig{IntrinsicSEDs} and \fig{Range}, but are left with strong features
at 7.7~\mum\ where the PAH features are over-estimated by the
host-galaxy component, indicating that this approach is less reliable
than our adopted approach.\footnote{We also get broadly similar
  results if we assume that 80\% or 99\% of the 100~\mum\ flux is due
  to the host-galaxy.}

\section{Comparison with previously defined AGN infrared SEDs and
  results from dusty torus models}
\label{Comparison}

We have produced a set of intrinsic AGN infrared SEDs by decomposing
the observed SEDS of local, AGN-dominated galaxies into well-defined
host-galaxy and intrinsic AGN components (see
\S\ref{Templates:AGN:FIR}).  The full variety of these intrinsic AGN
SEDs is shown in \fig{Range}.  In this section we explore how these
intrinsic SEDs compare against other commonly assumed AGN SEDs,
including those produced by radiative transfer models of the dust
surrounding the AGN.  For simplicity, the comparisons made in this
section are against the mean average and range of {\it all} the
intrinsic SEDs in our sample (i.e., we do not differentiate between
the high and low luminosity AGNs in our sample; see previous section).

\subsection{NGC~1068, Mrk~231}
\label{Comparison:NGC1068}

Of all known local AGNs, NGC~1068 and Mrk~231 are commonly used to
characterise the infrared SEDs of AGNs.  In \fig{CompareObs} we
illustrate how our range of intrinsic AGN SEDs compare with these
canonical AGNs.\footnote{The NGC~1068 and Mrk~231 SEDs shown here
  represent the infrared emission produced by the entire galaxy (i.e.,
  including emission from the host galaxy).  We note that, whilst
  mid-infrared flux density measurements of the resolved AGN core of
  NGC~1068 (e.g., \citealt{Gandhi09, Prieto10}) are now available, we
  choose to include emission from the host galaxy to allow easier
  comparison with previous studies which have, in general, done the
  same.} When normalised to 19~\micron, the SEDs of Mrk~231 and
NGC~1068 are, respectively, $\sim$15 and $\sim$8 times higher at
100~\micron\ than the average SED calculated from our AGN templates.
Assuming that the average intrinsic AGN emission continues to fall as
a modified black body to 1000~\micron, the Mrk~231 and NGC~1068 SEDs
contain roughly 1.5-3 times the power at infrared wavelengths than the
average intrinsic AGN SED (when all are normalised at 19~\micron).
This excess power is likely the result of host galaxy contamination in
these two cases, as suggested by \cite{Telesco84, Downes98} and
\cite{LeFloch01}.  Indeed, fits to the infrared SEDs of Mrk~231 and
NGC~1068 using the approach outlined in \S\ref{Templates:AGN:FIR} are
consistent with them being host-galaxy dominated at FIR wavelengths.
Finally, we note that at wavelengths $\lesssim20$~\micron, where the
host-galaxy contributions are small, the SEDs of NGC~1068 and Mrk~231
are largely consistent with our intrinsic AGN SEDs.

\subsection{Infrared Quasar SEDs}
\label{Comparison:QSO}

There have been a number of attempts to constrain the average quasar
SED, although only a handful have included coverage to FIR wavelengths
(e.g. \citealt{Elvis94, Richards06, Netzer07}).  One of the most
prominent, recent studies is that of \cite{Richards06}, which combined
photometric data covering the radio to X-ray regimes for a sample of
optically-selected, broad-line quasars.  In \fig{CompareObs} we
compare the average infrared SED of all quasars from that study to the
AGN templates derived here.  The average quasar SEDs from
\cite{Richards06} turns over at approximately the same FIR wavelengths
as the average of our intrinsic AGN SEDs, although we note that
host-galaxy contamination at FIR wavelengths has not been removed from
this average quasar SED.  This host-galaxy contamination will tend to
push the position of the turn-over to longer wavelengths.  The
\citeauthor{Richards06} quasar SED is flatter than the average
intrinsic SED of more moderate luminosity AGNs, which could be due to
the increased quasar luminosities heating the surrounding dust to
higher temperatures compared to more moderate luminosity AGNs.
Despite being flatter at short wavelengths, the \cite{Richards06} SED
contains only $\sim$10 per cent more flux over 8-1000~\mum\ than the
average intrinsic SED when both are normalised to 19~\mum\ and
extrapolated to longer wavelengths assuming a modified blackbody SED.

Recently, \cite{Netzer07} removed the host-galaxy component from the
average infrared SED of a sample of PG-quasars to produce an average,
intrinsic quasar infrared SED.  This SED is also investigated in
\fig{CompareObs}.  When both are normalised at 19~\mum\, the average
intrinsic quasar SED emits more strongly at $\lambda<$19~\mum\ than
the average intrinsic SED of more moderate luminosity AGNs calculated
above.  However, this situation is reversed at $\lambda>$19~\mum, with
the former falling more rapidly at longer wavelengths.  This
strengthens our findings described in \S\ref{Templates:AGN:FIR}, where
we report a similar trend between the higher (i.e., ${\rm
  log}($\lx$)>$42.9) and lower (i.e., ${\rm log}($\lx$)<$42.9)
luminosity AGNs within our sample.  We note that the intrinsic quasar
falls even more rapidly than the average intrinsic SED of the higher
luminosity AGNs in our sample.  The intrinsic quasar SED has stronger
silicate emission features than the average intrinsic SED of more
moderate luminosity AGNs, although this may be due to the
\cite{Netzer07} being comprised of Type 1 AGNs which tend to have
stronger silicate emission features (see \S\ref{Templates:AGN:MIR}).
As we suggest in \S\ref{Templates:AGN:FIR} the increased flux at short
wavelengths and relative deficit of flux at longer wavelengths in the
average quasar SED may be related to the higher luminosities of
quasars, with more luminous objects capable of heating more dust to
higher temperatures.  However, we can only claim this in an average
sense as the average quasar SED is consistent with the spread of
intrinsic infrared SEDs extracted from the 11 AGN-dominated sources
above.  The four intrinsic AGN SEDs that bear the closest similarity
to the average quasar intrinsic SED are 3C120, IC4239A, Mrk~509 and
NGC~4507; three are type 1 AGNs and one is a type 2 AGN, respectively.
All four of these SEDs have intrinsic 2-10~keV luminosities greater
than the median of our sample, (i.e., ${\rm log}($\lx$)>$42.9).

\begin{figure}
\includegraphics[width=85mm]{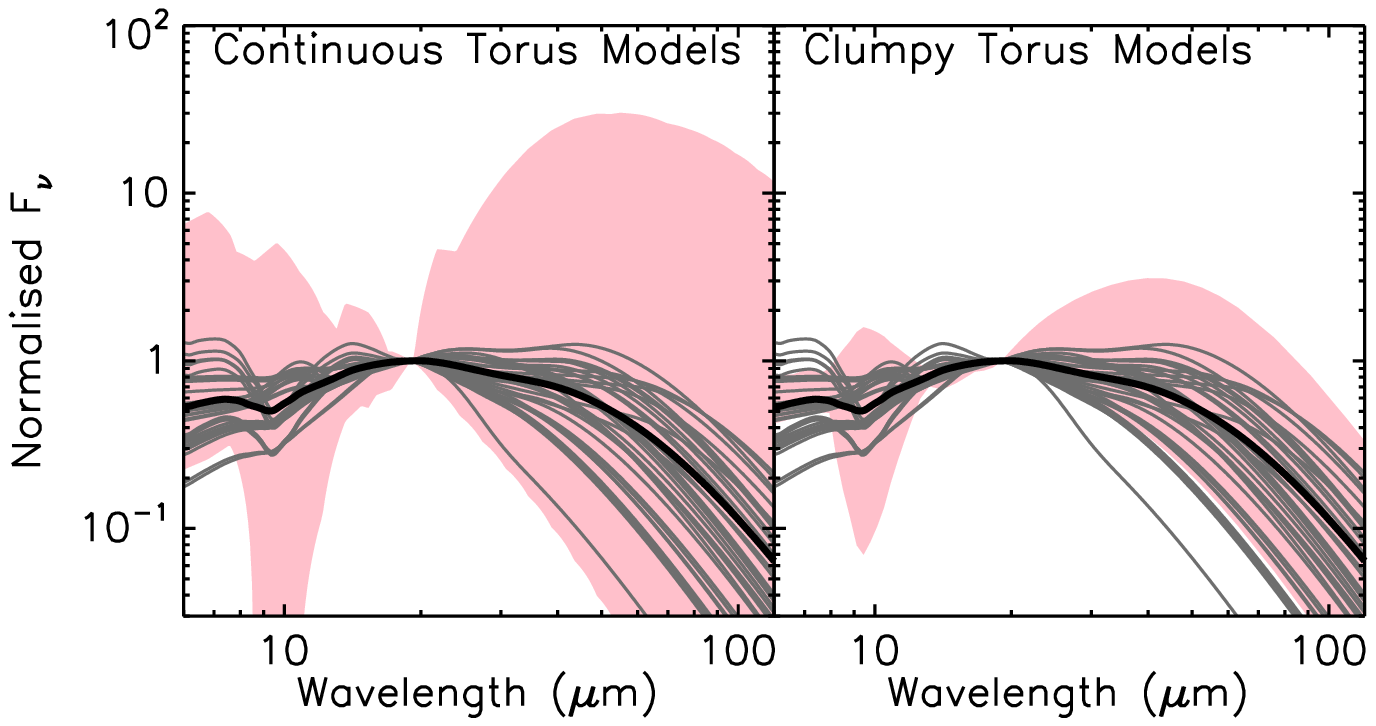}
\caption{The range of the intrinsic AGN infrared SEDs (as shown in
  \fig{CompareObs}, left panel) plotted against the range of SEDs
  predicted by continuous (left-hand plot) and clumpy (right-hand
  plot) torus models described in \protect \cite{Fritz06} and \protect
  \cite{Schartmann08}, respectively (shaded regions in the left and
  right panels, respectively). Each set of SEDs has been normalised at
  19~\micron.  In general, a clumpy torus model provides a better
  representation of the intrinsic AGN SED than the continuous torus
  models which over-predict the range of intrinsic AGN infrared SEDs by
  at least one order of magnitude at $\sim$60~\micron\ with this
  normalisation.  However, none of the clumpy torus models predict
  6-20~\micron\ SEDs as steep as some of those in our AGN-dominated
  sample (e.g., Mrk~3)}
\label{CompareMod}
\end{figure}

For comparison, we fit the \cite{Netzer07} intrinsic quasar SED using
\dcmb\ (see appendix \ref{Tests:Outline}), obtaining:

\begin{equation}
F_\nu \propto \left\{
\begin{array}{rcl}
\lambda^{1.2} & \mbox{at} & 6~\umu{\rm m} < \lambda < 20~\umu{\rm m}\\
\nu^{1.5}F_\nu^{\rm BB} & \mbox{at} & \lambda > 20~\umu{\rm m} \\
\end{array}\right.
\end{equation}
\noindent
where the symbols are the same as those defined in
\S\ref{Templates:AGN:FIR}.  These parameters confirm the shallower SED
at MIR wavelengths and shorter-wavelength turnover of the
\cite{Netzer07} intrinsic quasar SED, compared to the average AGN SED
defined in this study.  When we extrapolate the \cite{Netzer07}
intrinsic quasar SED to longer wavelengths as a modified blackbody and
normalise both SEDs at 19~\micron, then we find that the total power
radiated at infrared wavelengths is $\sim$80 per cent that of the
average intrinsic infrared SED of typical AGNs calculated here.

\subsection{Radiative transfer (dusty torus) models}
\label{Comparison:Torus}

Finally, we compare the range of intrinsic, infrared AGN SED templates
with predictions from radiative transfer models of the dust
surrounding the active nucleus, which is thought to be the main source
of infrared emission from AGNs (although there is some evidence to
suggest that infrared emission from AGNs could also be produced in
more extended, dusty regions; e.g. \citealt{Schweitzer08, Mor09}).  We
compare our templates with the SEDs produced by both (a) continuous
and (b) clumpy distributions of dust (specifically, the models
described in \citealt{Fritz06} and \citealt{Schartmann08},
respectively). The range of input parameters of the former (i.e.,
continuous) model are given in table 1 of \cite{Fritz06}.  The input
parameters of the clumpy torus model is presented in table 1 of
\cite{Schartmann08}; we used the SEDs produced by varying both
orientation angle (see their fig. 4) and dust mass (see their
fig. 10).  The SEDs predicted by these torus models are shown in
\fig{CompareMod}, together with the average and range of intrinsic
SEDs defined in this study.

\begin{figure}
\includegraphics[width=85mm]{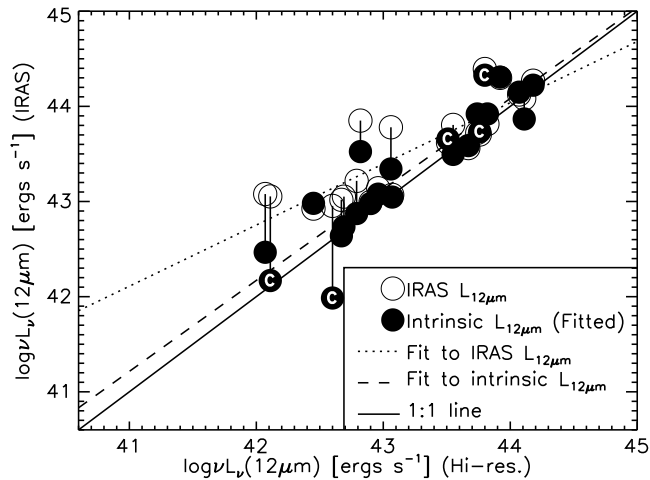}
\caption{The intrinsic 12~\micron\ AGN luminosities derived from our
  fits to the \IRAS\ photometry plotted against the core 12~\micron\
  luminosities obtained using high spatial resolution (i.e.,
  sub-arcsecond) mid-infrared observations of 23 nearby AGNs, taken
  from \protect \cite{Horst08} and \protect \cite{Gandhi09}.  In the
  latter, the host galaxy contribution is spatially resolved,
  providing an uncontaminated measure of the intrinsic AGN infrared
  luminosity at 12~\micron.  After fitting the large-aperture (i.e.,
  arcminute) \IRAS\ photometry and excluding any host galaxy
  contamination we reproduce, on average, the intrinsic luminosities
  derived from high spatial resolution observation to within a factor
  of $\sim$2, compared to a factor of $\sim$4 when the 12~\mum\ IRAS
  photometry is not corrected for host-galaxy contamination.  Points
  associated with Compton-thick AGNs have been marked with a ``C'' and
  show a similar level of scatter as the Compton thin sources.}
\label{Core}
\end{figure}

The continuous dust-distribution models produce a much broader range
of SED shapes than the intrinsic AGN infrared SEDs defined here.  For
example, when normalised at 19~\micron\ these models predict a range
of flux densities spanning almost three orders of magnitude at
$60$~\micron, compared to a spread of (at most) two orders of
magnitude for the intrinsic SEDs defined in this study.  We note that
the model SEDs that match the range of intrinsic AGN SEDs do not
correspond to a particular region of input parameter space (i.e.,
torus opening angle, optical depths, inner and outer torus radii,
radial dust density distribution, angular dust density distribution,
SED of incident radiation).  Therefore, our intrinsic AGN SEDs are
unable to directly constrain acceptable input parameters of these
continuous dust-distribution models.

The SEDs produced by the clumpy torus models provide a closer match to
the observed intrinsic AGN infrared SEDs, although they too
over-predict the range of FIR fluxes spanned by our intrinsic SEDs and
tend to peak (in $\nu F_{\nu}$) at longer wavelengths. The closer
match to the clumpy torus models is consistent with other evidence in
support for such models (e.g., \citealt{Horst06, Gandhi09, Ramos09}).
However, none of the clumpy torus models described in
\cite{Schartmann08} predict 6-20~\micron\ SEDs as steep as some of
those in our AGN-dominated sample (e.g., Mrk~3).  As a consequence, at
this normalisation the clumpy torus models over-predict the emission of
many of our sources at $\lambda\lesssim19$~\mum.

\section{Testing the AGN and host-galaxy templates}
\label{Tests}

A principal motivation behind defining the intrinsic infrared SEDs of
moderate luminosity AGNs is to use these as templates to fit the
infrared SEDs of composite galaxies and calculate the AGN and host
galaxy contributions to their total infrared output.  This is
especially useful in the majority of cases when only broad-band
infrared photometry measurements are available to constrain the
infrared SED.  In this section we explore whether fits to the
broad-band photometry incorporating the host-galaxy and AGN templates
defined here can reproduce the intrinsic AGN luminosities and AGN
contributions derived using other, independent approaches (i.e., high
resolution imaging of nearby AGNs and emission line diagnostics).  To
perform these fits, we have developed an {\sc idl} procedure, \dcmb,
which we describe in appendix \ref{Tests:Outline} and is publicly
available online at http://sites.google.com/site/decompir.

\begin{figure}
\includegraphics[width=85mm]{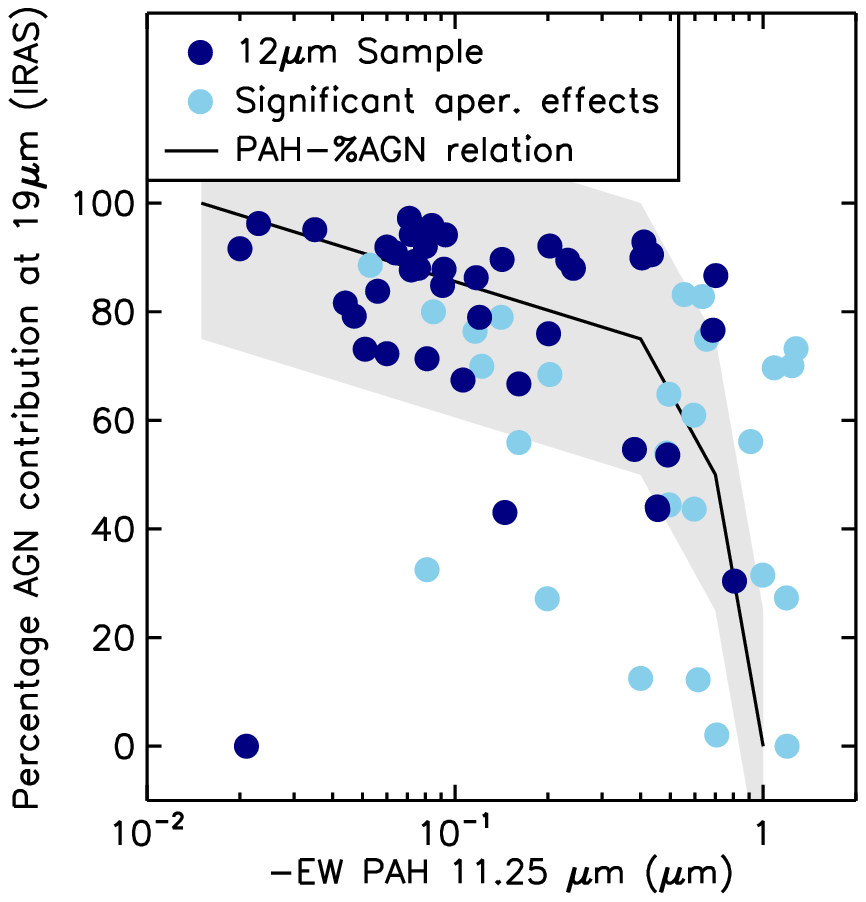}
\caption{\ewpah\ plotted against the AGN contribution at 19~\micron\
  derived from our fits to \IRAS\ photometry of 78 sources in the
  12~\micron\ sample of Seyfert galaxies (\protect \citealt{Rush93,
    Tommasin08, Tommasin10}).  Sources where the \IRAS\ 12~\micron\
  photometry measurement is at least 2 times higher than the flux at
  the same wavelength in the high resolution IRS spectrum in which the
  \ewpah\ is measured have been indicated.  Such cases are likely to
  suffer from significant aperture effects which will typically lead
  to more of the host galaxy flux being included in the \IRAS\
  photometry measurements.  Also shown in this plot is the \ewpah\ --
  percent AGN contribution relation taken from \protect
  \cite{Tommasin10} (black line) with a $\pm25$ per cent margin shown
  in grey.  We note a general agreement between these two measures of
  the host-galaxy contribution, with $\sim76$ per cent (i.e., 59/78)
  of AGN contributions measured from the photometry fits lying within
  25 per cent of that expected using \ewpah.}
\label{PAHIRASPerc}
\end{figure}

\subsection{Comparison with high spatial resolution observations of
  local Seyferts.}
\label{Tests:HiRes}

A number of recent studies (e.g. \citealt{Krabbe01, Horst06, Horst08,
  Gandhi09}) have shown that the intrinsic infrared luminosities of
nearby AGNs can be reliably measured from high spatial resolution
observations (i.e., $\sim0.5$\arcsec, typically probing physical
scales of $\lesssim100$~pc).  At such high resolutions, host galaxy
contamination is minimised, leaving only the emission intrinsic to the
active nucleus.  However, such analyses are only viable for a small
number of systems that are nearby enough that the central regions can
be sufficiently well resolved.  On the other hand, if we can use SED
decomposition to exclude the host galaxy contribution to the
broad-band SEDs, then we could potentially measure the intrinsic
12~\micron\ luminosities for all AGNs with well sampled infrared SEDs.
Here, we demonstrate that the SED decomposition procedure outlined in
this study can, indeed, be used to measure the intrinsic AGN
12~\micron\ flux.

We use \dcmb\ (see appendix \ref{Tests:Outline}) to fit the four-band
\IRAS\ photometries of 23 AGNs with core 12~\micron\ luminosities
reported in \cite{Horst08} or \cite{Gandhi09} and X-ray luminosities
within the range spanned by the AGNs used to define our intrinsic AGN
SEDs (i.e., \lx$=10^{42}-10^{44}$~\ergs).  As we are limited to only
four independent flux density measurements we only allow the
normalisations of the AGN and host galaxy components to vary when
fitting the observed SEDs.  All other parameters are fixed to their
average values derived from the full sample of 11 AGN dominated
galaxies described above (also see appendix \ref{Tests:Outline}), but
we note that our main results do not change if we use the average SED
of either the high (i.e., log(\lx)$>$42.9) or low (i.e.,
log(\lx)$<$42.9) X-ray luminosity AGNs in our sample.  Each SED was
fit five times, once for each of our host-galaxy templates (i.e., as
recommended in appendix \ref{Tests:Outline}).  To calculate the
intrinsic 12~\micron\ AGN flux, we integrate the intrinsic AGN
component derived from the best fitting solution of these five fitted
SEDs (i.e., that with the lowest associated $\chi^2$ value) over a
narrow (i.e., 1~\micron) top-hat passband centred at 12~\micron, which
is a similar response function as those filters used for the
observations described in \cite{Horst08} and \cite{Gandhi09}.

In \fig{Core} we plot the intrinsic 12~\micron\ flux derived from our
best fitting SEDs against those obtained using high spatial resolution
observations as reported in \cite{Horst08} and \cite{Gandhi09}.  We
find that the intrinsic luminosities derived using our SED fits are
well matched to those obtained using high resolution observations of
the AGN core.  On average, the intrinsic 12~\micron\ AGN luminosity
derived from our fits lie within a factor of two of the core
luminosity at this wavelength, compared to within a factor of four
when the 12~\mum\ IRAS photometry is not corrected for host-galaxy
contamination.  A linear regression to these points gives:

\begin{equation}
\begin{array}{ll}
{\rm log}\left( \frac{L_{\rm 12}^{\rm Fit}}{10^{43} {\rm erg~s^{-1}}}\right) = &
(0.13 \pm 0.07) + \\
&(0.96\pm0.09)
{\rm log}\left( \frac{L_{12}^{\rm HiRes}}{10^{43}{\rm erg~s^{-1}}}\right)\\
\end{array}
\end{equation}

\noindent
which has a slope consistent with a 1:1 relationship.  For
comparison, we also plot the 12~\micron\ luminosities calculated
directly from the 12~\micron\ \IRAS\ photometry, noting that these
often overestimate the intrinsic 12~\micron\ luminosities, sometimes
by more than a factor of 5.  Finally, we note that the five
Compton-thick sources in this sample (i.e., NGC~1068, IC~3639,
Swift~J0601.9-8636, NGC~5728 and NGC3281; \citealt{DellaCeca08})
display a similar degree of scatter as the full sample.  This
demonstrates the power of using infrared SED fitting to identify even
heavily obscured AGNs.

\subsection{Comparison with results derived from emission line
  diagnostics}
\label{Tests:PAH}

A number of infrared emission line diagnostics have recently been
identified as providing reliable proxy measures of the AGN and
host-galaxy contributions to the infrared and bolometric output of
composite galaxies (e.g. \citealt{Genzel98, Tommasin08, Tommasin10,
  Goulding09}).  A key test for our fits is whether they can broadly
reproduce the intrinsic AGN luminosities derived from these
diagnostics.

We use \dcmb\ to fit the \IRAS\ photometries of 78 galaxies from the
{\it IRAS} 12~\mum\ sample for which accurate emission line fluxes,
measured from high resolution IRS spectra, are available (see
\citealt{Tommasin08, Tommasin10} for a complete list of all the AGNs
used for this section of our study, together with their measured
emission line fluxes).  Again, as we are dealing with only four
photometry measurements, we only allow the normalisations of the AGN
and host-galaxy components to vary when fitting the observed data.
All other parameters are fixed to their average values derived from
the full sample of 11 AGN-dominated galaxies (see
\S\ref{Templates:AGN:FIR}; again, we note that our results do not
change if we use the average SEDs derived from the high (i.e.,
log(\lx)$>$42.9) or low (i.e., log(\lx)$<$42.9) X-ray luminosity AGNs
in our sample).  The \IRAS\ photometry of each of the 78 objects in
our sample are fitted five times, each time using a different choice
of host-galaxy template (defined in \S\ref{Templates:SB}).  Here, we
report the AGN contribution to the 19~\micron\ infrared emission
measured from the best fitting of these five independent fits (i.e.,
that with the lowest associated $\chi^2$ value).

In \fig{PAHIRASPerc} we plot the AGN contribution at 19~\micron\
derived from our fits to the \IRAS\ photometry against \ewpah\
measured from high resolution IRS spectra; see \cite{Tommasin08,
  Tommasin10}.  We find that the AGN contribution measured from our
fits to the \IRAS\ photometry are in good general agreement with those
based on \ewpah, especially at high levels of AGN contribution (i.e.,
$\gtrsim50$ per cent).  Of the entire sample of 78 AGNs, $\sim$76 per
cent (i.e., 59/78) have AGN contributions measured from our broad band
SED fits that lie within $\pm$25 per cent of those derived from
\ewpah.  At lower AGN contributions where the intrinsic AGN component
is more difficult to constrain, our fits typically estimate the AGN
contribution to within $\sim50$ per cent of that derived from \ewpah.
We note, however, that at such low AGN contributions even emission
line diagnostics have considerable difficulty in estimating the
intrinsic AGN contribution, as demonstrated by the significant amounts
of scatter in the \nev/\neii\ -- \ewpah\ correlation shown in fig. 4
of \cite{Tommasin10} (see also right-hand panel of fig. 8.,
\citealt{Goulding09})

\section{Application of our templates and infrared photometry fitting
  procedure}
\label{Application}

In the previous section, we quantified the accuracy of our SED fitting
routine by comparing the results from these fits to results obtained
by other well-established approaches In this section, we demonstrate
two additional applications of our AGN intrinsic AGN SEDs and fitting
approach: (a) defining the correction factors to convert 12~\micron\
and 2--10~keV luminosities to total AGN infrared luminosities (i.e.,
\liragn) and (b) measuring the intrinsic AGN luminosities of large
samples of composite galaxies for which only infrared photometry
measurements are available.

\subsection{$\nu L_\nu(12~\micron)$:\liragn\ and \lx:\liragn\ correction
  factors}
\label{Application:Bol}

\begin{figure}
\includegraphics[width=85mm]{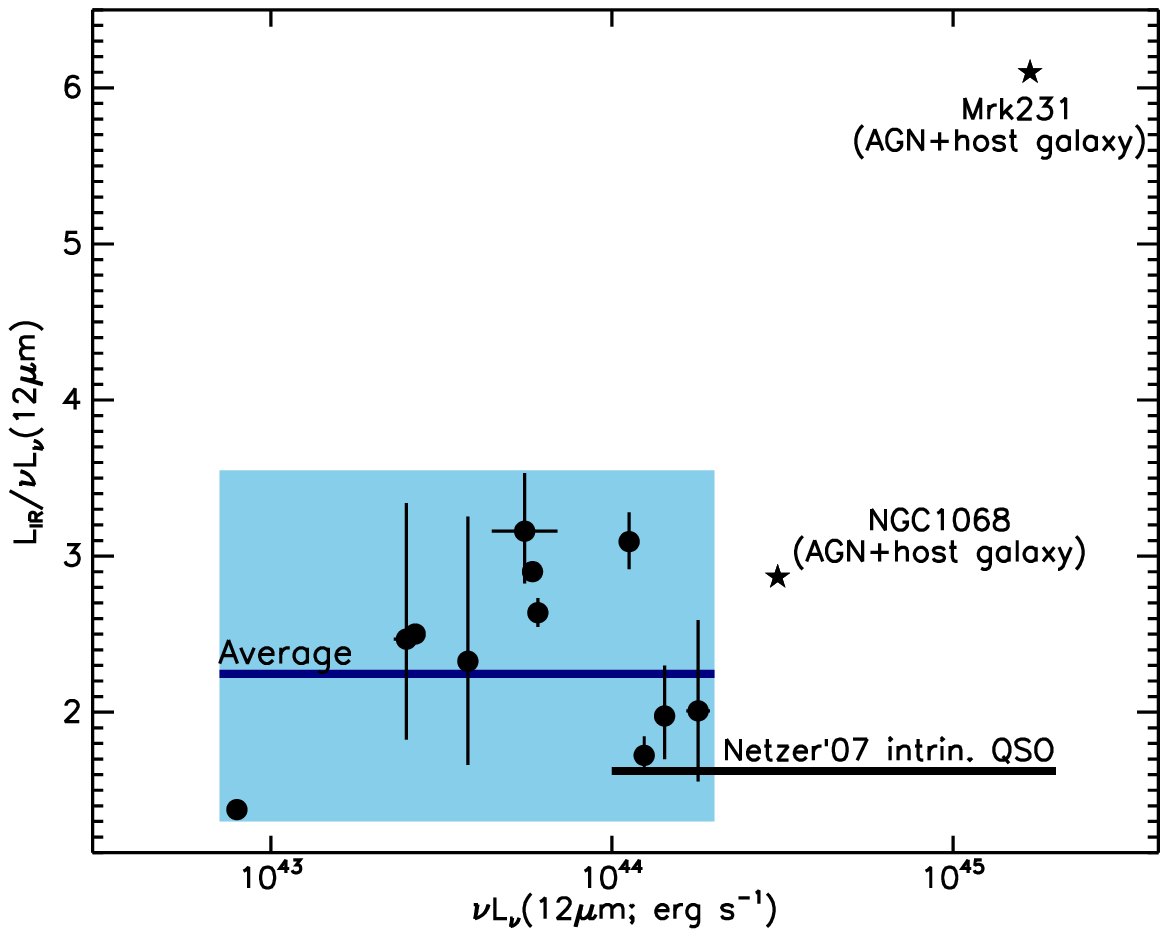}
\caption{The range of \liragn/\ltwelve\ ratios plotted against
  \ltwelve\ for the 11 AGNs that we use to define the range of
  intrinsic AGN infrared SEDs (filled circles).  We also show the
  \liragn/\ltwelve\ ratio for the average intrinsic AGN infrared SED
  defined in \S\ref{Templates}, the average intrinsic quasar infrared
  SED from \protect \cite{Netzer07} and the two archetypal AGNs,
  NGC~1068 and Mrk~231 (the average quasar SEDs of \protect
  \citealt{Elvis94} and \protect \citealt{Richards06} have
  approximately the same \liragn/\ltwelve\ ratios as the \protect
  \citealt{Netzer07} quasar SED; i.e., \liragn/\ltwelve\
  $\approx$1.60, 1.59 and 1.62, respectively). On average, the
  moderate luminosity AGNs considered in this study have higher
  intrinsic \liragn/\ltwelve\ ratios than more luminous AGNs (i.e.,
  quasars), although there is no clear correlation between \ltwelve\
  and \liragn/\ltwelve\ for the 11 AGNs in our sample.  We wish to
  point out that, as we have defined the average intrinsic AGN SED at
  all infrared wavelengths (i.e., 6--1000~\mum), an equivalent plot
  could be produced for any wavelength in this range (see
  \tab{SED_Table})}
\label{BolCorr}
\end{figure}

\begin{table*}
    \begin{minipage}{114mm}
      \caption{Average AGN and host-galaxy SEDs and the
        wavelength-dependent correction factors used to calculated
        \liragn and $L_{\rm IR}^{\rm QSO}$.}
    \begin{center}
    \input{table3.tex}
    \end{center}
    \label{SED_Table} {\sc Notes}: (1) Wavelength, (2)-(4) Flux
    densities of the average intrinsic AGN infrared SEDs defined in
    \S\ref{Templates}, normalised at 6~\micron.  Hi. Lum and Lo. Lum
    refer to the average SEDs of log(\lx)$>42.9$ and log(\lx)$<42.9$
    AGNs in our sample, respectively. (5)-(9) Flux densities of the
    five host galaxy templates defined in \S\ref{Templates:SB},
    normalised at 6~\micron, (10) Conversion factors to convert $\nu
    L_\nu^{\rm AGN}$ to total intrinsic AGN infrared luminosities
    (i.e., $L_{\rm IR}^{\rm AGN}$) at all wavelengths covered by our
    templates, based on the average AGN SED defined in
    \S\ref{Templates}. (11) The same as column 10, but for the average
    quasar template of \protect \cite{Netzer07}.  This table is
    available in its entirety at higher wavelength resolutions at
    http://sites/google.com/site/decompir.
\end{minipage}
\end{table*}

In cases where the infrared SED is either poorly constrained or
dominated by emission from the host galaxy, measuring the total
intrinsic AGN contribution can prove extremely difficult.
Furthermore, to accurately estimate key parameters such as
star-formation rates from infrared emission requires that any AGN
contribution be accounted for.  However, often, the only accurate way
to measure the levels of AGN activity in composite galaxies is via
X-ray or high resolution infrared observations.  Using the intrinsic
AGN infrared SEDs defined here coupled with the correlation between
the intrinsic 12~\micron\ luminosity and the 2--10~keV luminosity
defined in \cite{Gandhi09}, we can now define a correction factor to
convert 2--10~keV luminosities to total intrinsic AGN infrared
luminosities.  This correction factor enables any AGN contribution to
the infrared output of galaxies for which, for example, only X-rays or
monochromatic infrared observations provide a measure of the intrinsic
AGN luminosity.

In \fig{BolCorr} we show the range of \liragn/$\nu L_{\nu}(12~\umu
{\rm m})$ ratios for all 11 AGNs for which we measure the intrinsic
AGN infrared SED, plotted against $\nu L_{\nu}(12~\umu {\rm m})$.
Here, we have assumed that the intrinsic AGN infrared SED continues as
a modified black body to 1000~\micron.  These corrections therefore
only apply to the thermal component of the intrinsic AGN SED as we do
not consider any emission from non-thermal (e.g., synchrotron)
processes.  We choose $\nu L_{\nu}(12~\umu {\rm m})$ as a reference
point as it has been shown to be strongly correlated with the
intrinsic luminosity of the AGN (e.g., \citealt{Gandhi09}).  However,
as we have defined the intrinsic AGN SED at 6-1000~\micron, an
equivalent figure/analyses could be made/performed using any wavelength
in this range as a reference point (see \tab{SED_Table}).  The
\liragn/$\nu L_{\nu}(12~\umu {\rm m})$ ratios displayed by these 11
AGNs span the range $\sim$1.3 to $\sim$3.5, with an average ratio of
$\sim2.2$ (derived from the average SED defined in \S\ref{Templates}).
We do not see any correlation between \liragn/$\nu L_{\nu}(12~\umu
{\rm m})$ and $\nu L_{\nu}(12~\umu {\rm m})$ for our sample.  The
average \liragn/$\nu L_{\nu}(12~\umu {\rm m})$ ratio for quasar SEDs
defined in \cite{Netzer07} is slightly lower (i.e., $\sim1.6$) than
that of the lower luminosity AGNs considered here, although this
difference is not significant based on the range of ratios spanned by
our sample of intrinsic AGN SEDs.  However, we note that the at other
wavelengths the \cite{Netzer07} quasar bolometric correction differs
considerably from the average bolometric correction of more modest
luminosity AGNs defined here (see \tab{SED_Table}).  We also show in
\fig{BolCorr} the \lir/$\nu L_{\nu}(12~\umu {\rm m})$ ratios for
NGC~1068 and Mrk~231, the latter of which is significantly higher than
the range of intrinsic ratios due to the additional host-galaxy
components in these SEDs (see \S\ref{Comparison:NGC1068}).

Using the $\nu L_{\nu}(12~\umu {\rm m})$:\lx\ correlation defined in
\cite{Gandhi09}, we convert our \liragn/$\nu L_{\nu}(12~\umu {\rm m})$
ratios into \liragn/\lx\ correction factors.  Assuming \liragn/$\nu
L_{\nu}(12~\umu {\rm m})=2.2\pm1.3$ (to cover the full range of ratios
displayed by our sample) we obtain:

\begin{equation}
\begin{array}{ll}
{\rm log}\left( \frac{L_{\rm IR}^{\rm AGN}}{10^{43} {\rm erg~s^{-1}}}\right) = &
(0.53 \pm 0.26) + \\
&(1.11\pm0.07)
{\rm log}\left( \frac{L_{\rm 2-10keV}}{10^{43}{\rm erg~s^{-1}}}\right)\\
\end{array}
\label{Eqn:Bol}
\end{equation}

For a \lx$=10^{43}$~\ergs\ AGN, typical of the range of \lx\
considered here, equation \ref{Eqn:Bol} gives \lx/\liragn$\sim$0.3.
This compares to a median ratio of \lx/\liragn$\gtrsim$0.04 for the
quasar sample of \cite{Elvis94}, although in this case there will be
some contribution to \liragn\ from the host galaxy that has not been
accounted for, hence the lower limit (see also \citealt{Alexander05};
we have assumed \lir$\sim1.3L_{\rm FIR}$).  We attribute the factor of
$\lesssim$7 difference between these ratios to both the
unaccounted-for host galaxy contribution and the different fractions
of the total bolometric luminosities of AGNs (i.e., $L_{\rm Bol}$)
that are emitted at X-ray energies by quasars and more modest AGNs
(see \citealt{Vasudevan07}).  For example, \lx$=10^{43}$~\ergs\ AGNs
emit approximately 5 per cent of their bolometric power in the
2--10~keV band, compared to just $\sim1$ per cent for a
\lx$=10^{45}$~\ergs\ quasar (see fig. 3 of \citealt{Vasudevan07};
i.e., changes in $\alpha_{\rm OX}$ with luminosity;
\citealt{Avni82,Wilkes94,Vignali03,Steffen06}).  Using these conversion
factors, we find that \lir/$L_{\rm Bol}$ are similar for both moderate
and high luminosity (i.e., quasar) AGNs (i.e., 0.17 and
$\lesssim$0.25, respectfully). \footnote{If we assume that
  \lir/$L_{\rm Bol}$ provides a measure of the covering fraction of
  the infrared emitting dust, then these numbers are at odds with
  other measures based on the relative numbers of Type 1 and Type 2
  AGNs in the local Universe (i.e., 0.5-0.8; e.g., \citealt{Huchra92,
    Rush93, Maiolino95, Maia03}).  However, clumpy torus models
  predict that many additional parameters (such as the temperature of
  the clouds, the inner and outer radii of the torus and the total
  number and distribution of dust clouds; \citealt{Nenkova02}) on top
  of the covering fraction play a role in determining \lir$/L_{\rm
    Bol}$.  Indeed, the clumpy torus models of \cite{Schartmann08}
  predict a range of \lir$/L_{\rm Bol}$ spanning $\sim0.15 - 0.3$
  despite their models all having a constant covering fraction of
  0.5.}

Finally, as we have defined the average intrinsic AGN infrared SED at
6~\micron\ to 100~\micron, we can use it to convert an intrinsic AGN
infrared flux obtained at any wavelength covered by our SED templates
into \liragn.  We publish these conversion factors in column 8 of
\tab{SED_Table} (which is available in its entirety online at
http://sites.google.com/site/decompir).

\subsection{Application of the SED fitting procedure to the Swift-BAT
  AGN sample}
\label{Application:Samples}

One of the principal applications of our SED fitting procedure is to
measure the AGN contribution to the infrared output of composite
galaxies.  Because our fitting procedure can work with sparsely
sampled photometry data (e.g. four-band \IRAS\ photometry) this opens
up the possibility of measuring the intrinsic infrared emission of
AGNs in large samples of galaxies.  Furthermore, as the intrinsic AGN
luminosity at 12~\micron\ is strongly correlated to \lx, we can use
these fits to estimate the intrinsic X-ray luminosities of those AGNs
that are too heavily obscured to be detected by the current suite of
X-ray observatories (e.g., \Chandra, {\it XMM-Newton} and {\it
  Suzaku}).  This is analogous to using high-resolution 12~\micron\
observations to measure the intrinsic AGN luminosities.  To
demonstrate this, we use the procedure described in appendix
\ref{Tests:Outline} to fit the 12~\micron, 25~\micron, 60~\micron\ and
100~\micron\ photometry of all of the X-ray AGNs in the \Swift-BAT
sample that were detected in all four \IRAS\ wavebands.

We use \dcmb\ (see appendix \ref{Tests:Outline}) to fit the \IRAS\
photometry of the 44 \Swift-BAT AGNs detected in all four infrared
bands.  As we are limited to only four independent flux density
  measurements we only allow the normalisations of the AGN and host
  galaxy components to vary.  We fix all of the other parameters to
  their average values derived from the full sample of 11
  AGN-dominated galaxies (see \S\ref{Templates:AGN:FIR} and appendix
  \ref{Tests:Outline}), but note that the results do not change if we
  instead use the average SEDs of the high (i.e., log(\lx)$>$42.9) or
  low (i.e., log(\lx)$<$42.9) X-ray luminosity AGNs in that sample.
  Each SED was fit five times, once for each of our host-galaxy
  templates (i.e., as recommended in appendix \ref{Tests:Outline}).
  We use fit with the smallest associated $\chi^2$ value to determine
  the AGN contribution to the infrared SED.

\begin{figure}
\includegraphics[width=85mm]{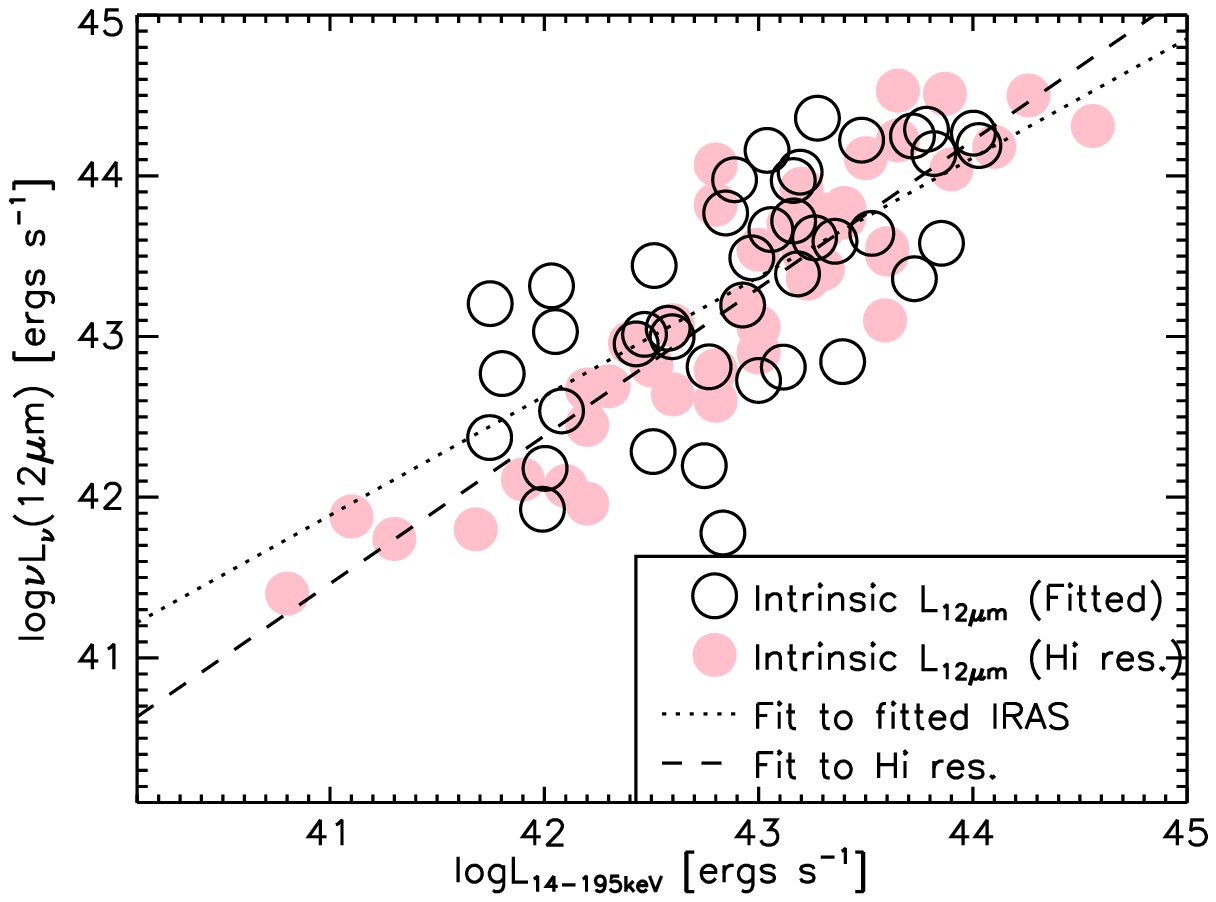}
\caption{\ltwelve\ derived from our {\sc dcmb} fits to the \IRAS\
  photometry of 44 AGNs from the \Swift-BAT sample plotted against the
  14--195~keV X-ray luminosities of these AGNs.  We see a clear
  correlation between these two measures of the intrinsic AGN
  luminosities. To give an impression of the intrinsic scatter in this
  relation, we also include 42 AGNs which have \ltwelve\ measured from
  high resolution observations.  For this latter sample, we convert
  \lx\ to $L_{\rm 14-195~keV}$ using the conversion $L_{\rm
    14-195~keV}$=2.5\lx.}
\label{L12LX}
\end{figure}

  In \fig{L12LX} we plot the intrinsic AGN 12~\micron\ luminosities of
  the 42 \Swift-BAT AGNs against their 14-195~keV X-ray luminosities
  (obtained from \citealt{Tueller08}).  Here, we use 14--195~keV
  luminosities, rather than 2--10~keV luminosities, as these energies
  are less susceptible to absorption and, as such, provide a more
  direct measure of the intrinsic AGN luminosity.  Each intrinsic AGN
  12~\micron\ luminosity is calculated by integrating the AGN
  component obtained from the best fitting SED over a narrow (i.e.,
  1~\micron) top-hat passband centred on 12~\micron.  We find that the
  intrinsic 12~\micron\ luminosity is strongly correlated with the
  14-195~keV luminosity.  For comparison, we also include in
  \fig{L12LX} the 42 AGNs presented in \cite{Gandhi09} that have
  published intrinsic 12~\micron\ luminosities measured from high
  resolution MIR observations.  For these, we have converted their
  intrinsic 2-10~\kev\ luminosities to 14-195~keV luminosities using a
  constant conversion factor of 2.5 (i.e., $L_{\rm
    14-195~keV}$=2.5\lx).  We note that there is an inherent
  uncertainty in this conversion factor meaning the true scatter of
  these $L_{\rm 14-195~keV}$ values will differ from that shown in
  \fig{L12LX}, although we expect any such differences to be
  comparatively small.\footnote{Assuming an X-ray spectral index of
    $\Gamma$=1.8, we get a factor of 2.7, for $\Gamma$=1.3 the factor
    is 10 and for $\Gamma$=2.3 it is 0.8, implying a factor of $\sim3$
    uncertainty for a typical range of $\Gamma$, although we note that
    other X-ray components (e.g. cold reflection, high energy
    turn-over) will also influence these factors.} The results derived
  from our SED fits show a similar level of scatter as those presented
  in \cite{Gandhi09}.  We note, however, that at low X-ray
  luminosities (i.e., $\lesssim10^{42}$~\ergs), our procedure tends to
  overestimate the intrinsic AGN luminosities compared to those
  obtained from high spatial resolution observations.  This is because
  the host galaxy tends to dominate over the AGN at lower
  luminosities, making it increasingly difficult to reliably constrain
  the AGN component.  Therefore caution should be taken when applying
  this kind of analyses to studies of lower-luminosity AGNs.  A linear
  regression to all \Swift-BAT AGNs considered here gives only modest
  errors, i.e.:

\begin{equation}
\begin{array}{ll}
{\rm log}\left( \frac{\nu L_{\rm \nu}^{\rm AGN}(12~\umu {\rm m})}{10^{43} {\rm erg~s^{-1}}}\right) = &
(0.37 \pm 0.08) + \\
&(0.74\pm0.13)
{\rm log}\left( \frac{L_{\rm 14-195keV}}{10^{43}{\rm erg~s^{-1}}}\right)\\
\label{Corr}
\end{array}
\end{equation}

\noindent
with a standard error (i.e., 1~$\sigma$ scatter) of $\sim0.5$ dex
(i.e., a factor of 3.2), compared to a standard error of $\sim0.36$
dex for the 42 comparison AGNs from \cite{Gandhi09} based on
high-resolution 12~\mum\ AGN luminosities.

The ability of our fitting procedure to derive the intrinsic
luminosities of AGNs from only sparsely sampled photometric infrared
data opens up the possibility of using such analyses to explore large
populations of AGNs detected in wavebands probed by the \IRAS, {\it
  ISO}, {\it AKARI}, {\it Spitzer} and {\it Herschel} telescopes.
Such studies will complement work at other wavelengths in exploring
the properties of both obscured and unobscured AGNs at all redshifts
probed by these facilities.  Applying our SED fitting procedure to
{\it Spitzer} and {\it Herschel} data will be the focus of a future
study.

\section{Summary}
\label{Summary}

Using a carefully selected sample of X-ray AGNs that show little or no
evidence of host-galaxy contamination in their \Spitzer-IRS
mid-infrared spectra, we have defined the range of intrinsic infrared
(i.e., 6-100~\micron) SEDs of typical, moderate luminosity (i.e.,
\lx$=10^{42}-10^{44}$~\ergs) AGNs found in the local Universe (i.e.,
$z<0.1$).  These intrinsic AGN SEDs can be used in conjunction with
host-galaxy templates to reproduce at least 85 per cent of the
6-35~\mum\ IRS spectra of composite galaxies to within $\pm10$ per
cent as well as their 60~\mum\ and 100~\mum\ photometries (see
\S\ref{Templates:AGN:FIR} and \fig{selectedAGNs}).  We outline a
procedure that uses the average intrinsic AGN infrared SED, in
conjunction with a set of host-galaxy templates, to fit the broad-band
infrared photometry of composite galaxies.  To test how accurately
this procedure can measure (a) intrinsic AGN luminosities and (b) the
AGN contribution to the total infrared output of composite galaxies,
we use it to fit the \IRAS\ photometry of two independent sample of
galaxies; the sample of core-resolved Seyferts described in
\cite{Horst08} and \cite{Gandhi09} and a subsample of the 12~\micron\
sample of Seyferts (e.g., \citealt{Rush93, Tommasin10}).  The results
from these tests compare favourably against other measures based on
high-spatial resolution mid-infrared observations and infrared
emission line diagnostics.

We summarise the main results of this study in the following points:

\begin{itemize}

\item The intrinsic SED of moderate luminosity (i.e.,
  \lx$=10^{42}-10^{44}$~\ergs) AGNs at 6-(15-60)~\micron\ (depending
  on turnover point, 15$<\lambda_{\rm BB}$/\micron$<$60) can be
  described as either a single or a broken power-law.  When we see a
  break, which typically lies between 15 and 20~\micron\ (mean:
  19~\micron), the power-law spectral indices shortward of the break
  span the range: $0.7<\alpha_1<2.7$ (mean: 1.7), where $F_\nu =
  \lambda^\alpha$.  Beyond this break, the intrinsic AGN SED always
  flattens, with spectral indices spanning the range $0<\alpha_2<1.5$
  (mean: 0.7). These ranges of spectral indices and and break
  positions are consistent with previous studies of the mid-infrared
  SEDs of AGNs (e.g., \citealt{Buchanan06, Wu09}). See
  \S\ref{Templates:AGN:MIR}.

\item All of the 11 AGN dominated infrared SEDs that we can reliably
  extrapolate to 100~\micron\ are consistent with being composed of a
  host-galaxy component plus an intrinsic AGN component.  All of the
  intrinsic AGN SEDs peak (in $\nu F_\nu$) between 15 and 60~\mum\
  before falling rapidly at longer wavelengths.  Despite this fall-off
  at long wavelengths, there are at least 3 AGNs in our sample whose
  observed infrared SEDs is dominated by the AGN even at 60~\mum. See
  \S\ref{Templates:AGN:FIR}

\item We find evidence that the shape of the intrinsic AGN infrared
  SEDs is related to AGN luminosity, with more luminous AGNs having
  bluer intrinsic SEDs (i.e., more luminous AGNs emit, relatively,
  more strongly at MIR wavelengths and less strongly at FIR
  wavelengths) .  This trend is seen both within our sample (see
  \S\ref{Templates:AGN:FIR}, \fig{Range} and \fig{CompareObs}) and
  when we compare our results with the intrinsic SEDs of more luminous
  quasars (see \S\ref{Comparison:QSO}).

\item The range of intrinsic AGN infrared SEDs reported here are more
  consistent with the results of clumpy, rather than continuous, dusty
  torus models.  These latter models, such as those described in
  \cite{Fritz06}, cover a much broader range of SED shapes than the
  intrinsic AGN infrared SEDs defined here.  However, we also note
  that even clumpy torus models have a tendency to over-predict the
  intrinsic infrared emission of AGNs at
  $\lambda\gtrsim30$~\mum. Furthermore, clumpy torus models do not
  produce 6-20~\mum\ as steep as that of a number of AGNs in our
  sample.  See \S\ref{Comparison:Torus}

\item Using the intrinsic AGN infrared SEDs defined in
  \S\ref{Templates} we define a set of correction factors to
  convert intrinsic AGN infrared or X-ray luminosities to total
  infrared luminosities (i.e., integrated 8-1000~\micron\
  luminosities).  These conversion factors can be used to estimate the
  AGN contribution to the total infrared output of galaxies when only
  a single (intrinsic) infrared or X-ray constraint is available.  See
  \S\ref{Application:Bol}.

\item Fits to the infrared SEDs of composite galaxies can be used to
  measure the intrinsic 12~\micron\ luminosities of AGNs to within a
  factor of $\sim$2 when only low-resolution, broad band photometry is
  available (\S\ref{Tests:HiRes}).  The same fits can be used to
  determine the intrinsic 2-10~keV X-ray luminosities to within a
  factor of $\sim$3 (\S\ref{Application:Samples}) and the fractional
  AGN contribution to the overall infrared output of a composite
  galaxy to within 25--50 per cent (depending on whether the galaxy is
  AGN dominated; \S\ref{Tests:PAH}).

\end{itemize}  

\section*{acknowledgements}
We would like to thank D. Elbaz and K. M. Dasyra for their useful
comments on the paper and M. Schartmann and E. Hatziminaoglou for
providing us with the results from their clumpy and continuous dusty
torus models, respectively.  Furthermore, we would like to thank the
referee, H. Netzer, for his comments, which significantly improved the
quality of the paper. We gratefully acknowledge support from the
Eurotalents Fellowship Program (JRM) the Leverhulme Trust (JRM; DMA),
the Royal Society (DMA), an STFC Studentship (ADG) and an STFC
Postdoctoral Fellowship (RCH).  This work is based (in part) on
observations made with the {\it Spitzer Space Telescope} and has made
use of the NASA/ IPAC Infrared Science Archive, which are both
operated by the Jet Propulsion Laboratory, California Institute of
Technology under a contract with NASA. Support for this work was
provided by NASA through an award issued by JPL/Caltech.

\bsp

\appendix

\section{Plots of fits to the IRS spectra and IRAS photometry}
 \label{SEDFits}
 In \fig{selectedAGNs} we present the observed SEDs of our
   sample of 11 AGNs that we use to defined the intrinsic infrared AGN
   SED.  We have included in each of these plots the fits to these
   SEDs produced using our five host-galaxy templates (see
   \S\ref{Templates:SB}) and an intrinsic AGN component (described in
   \S\ref{Templates:AGN:FIR}).  Embedded in each plot is a panel
   showing a zoomed-in image of the fit to the 6-35~\mum\ IRS spectrum
   and the residuals produced by subtracting the fit from the observed
   data.

\begin{figure*}
\includegraphics[width=160mm]{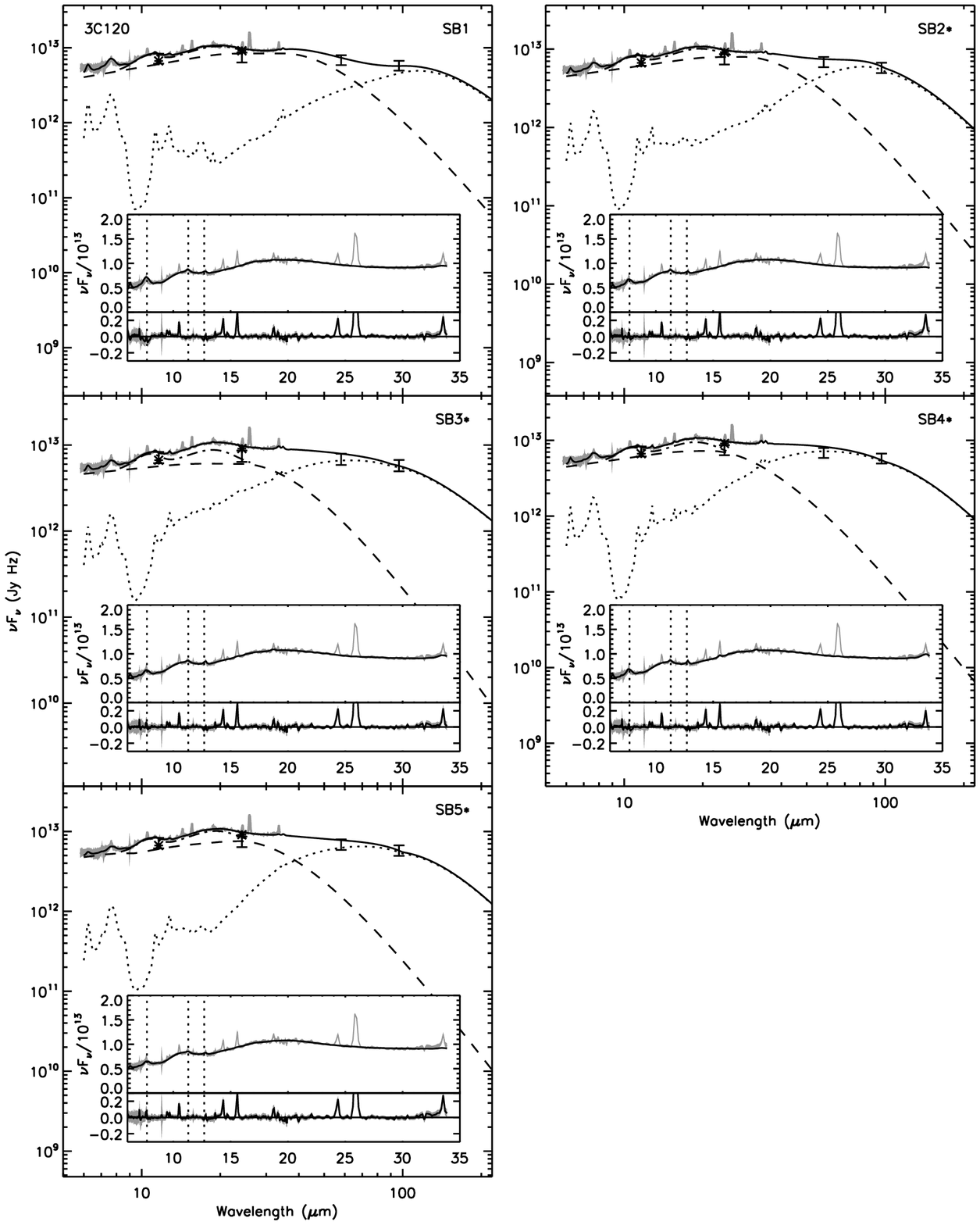}
\caption{On this and the following 10 pages we present the 11
  AGN-dominated SEDs that we use to extend the intrinsic AGN infrared
  SEDs to 100~\micron.  The five panels are used to show the results
  of fitting the observed SEDs with models containing the five
  separate host-galaxy components described in \S\ref{Templates:SB}.
  In each panel the IRS spectrum ($\pm5\sigma$) is shown as grey
  shading.  In each case, the model SED fit is represented as a solid
  black line.  The AGN component is shown as a dashed line and the SB
  component is shown as dotted line.  Where present, the silicate
  emission at $\sim$10~\mum\ and $\sim$18~\mum\ is shown as a dash-dot
  line on top of the AGN component (see \S\ref{Templates:AGN:FIR} for
  further details on the fitting procedure).  A zoomed-in image of the
  fit to the 6-35~\micron\ portion of the IRS data is shown inset,
  together with the residuals of the fit (the positions of the PAH
  features at 7.7~\micron, 11.25~\micron\ and 12.8~\micron\ are
  highlighted with vertical dotted lines).  We use \IRAS\ 60~\micron\
  and 100~\micron\ photometry to constrain the emission at
  far-infrared wavelengths.  We have placed an asterisk next to the
  starburst name of each fit that is considered good based on the
  criteria outlined in \S\ref{Templates:AGN:FIR}.}
\label{selectedAGNs}
\end{figure*}

\begin{figure*}
\includegraphics[width=160mm]{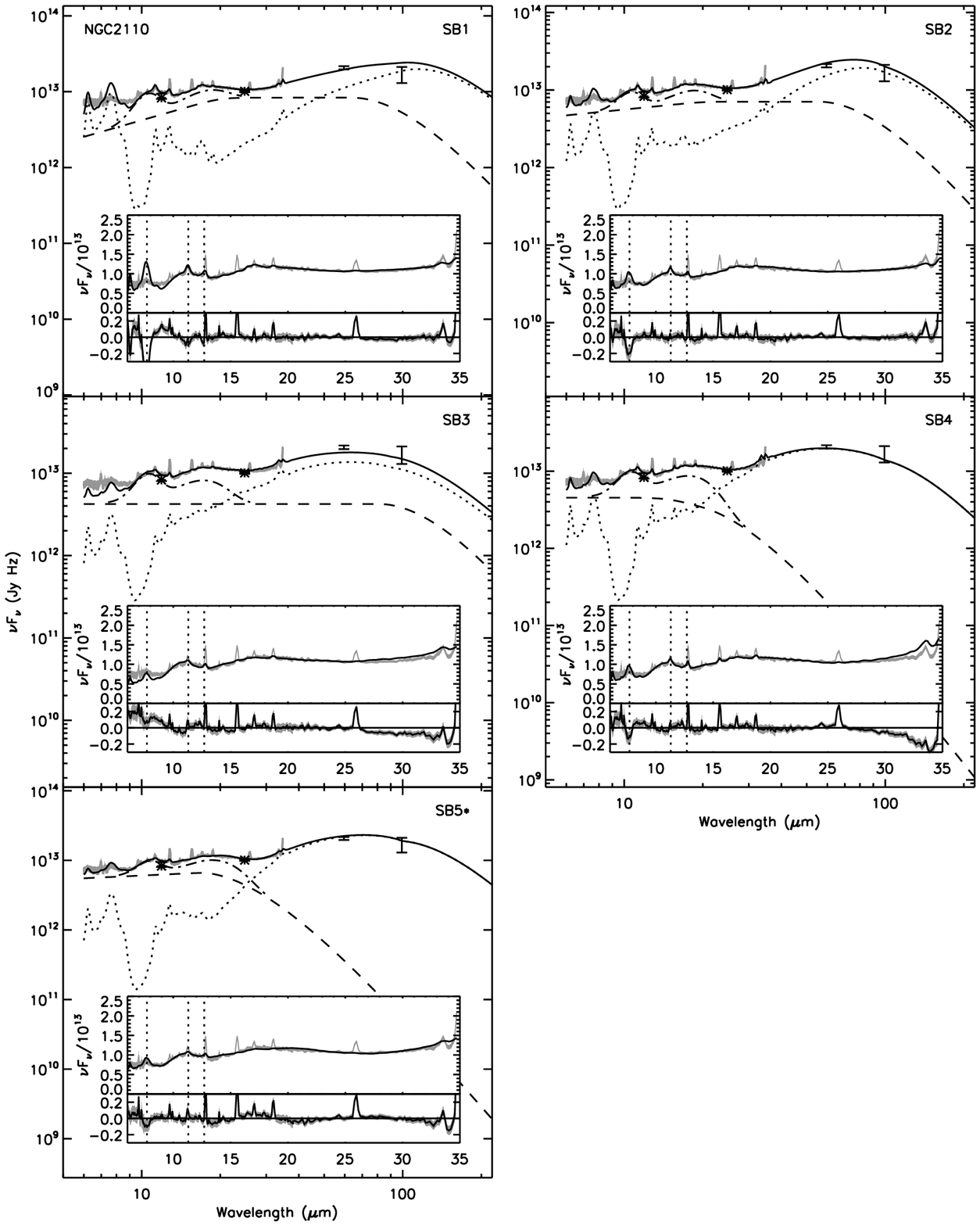}
\contcaption{}
\end{figure*}

\begin{figure*}
\includegraphics[width=160mm]{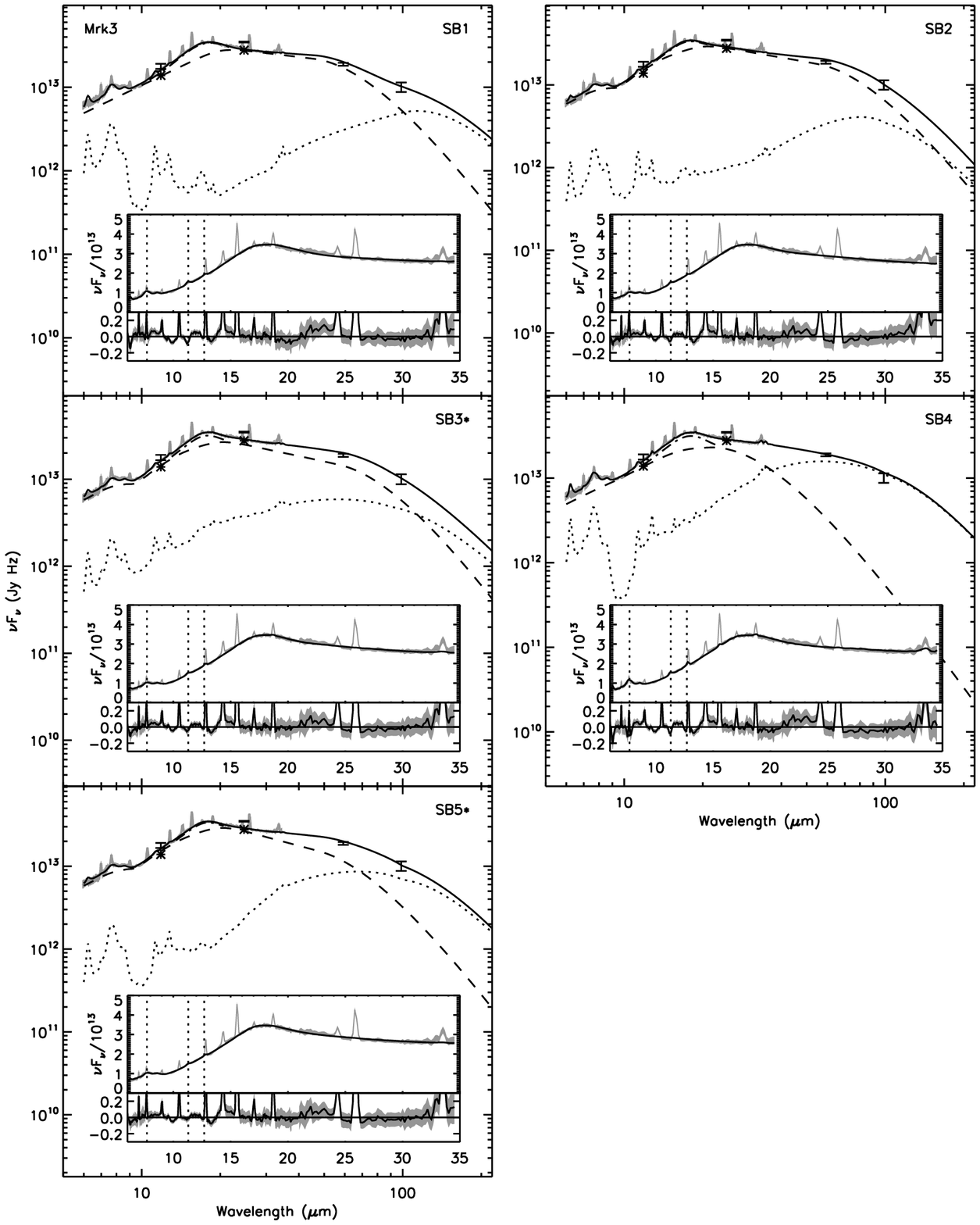}
\contcaption{}
\end{figure*}

\begin{figure*}
\includegraphics[width=160mm]{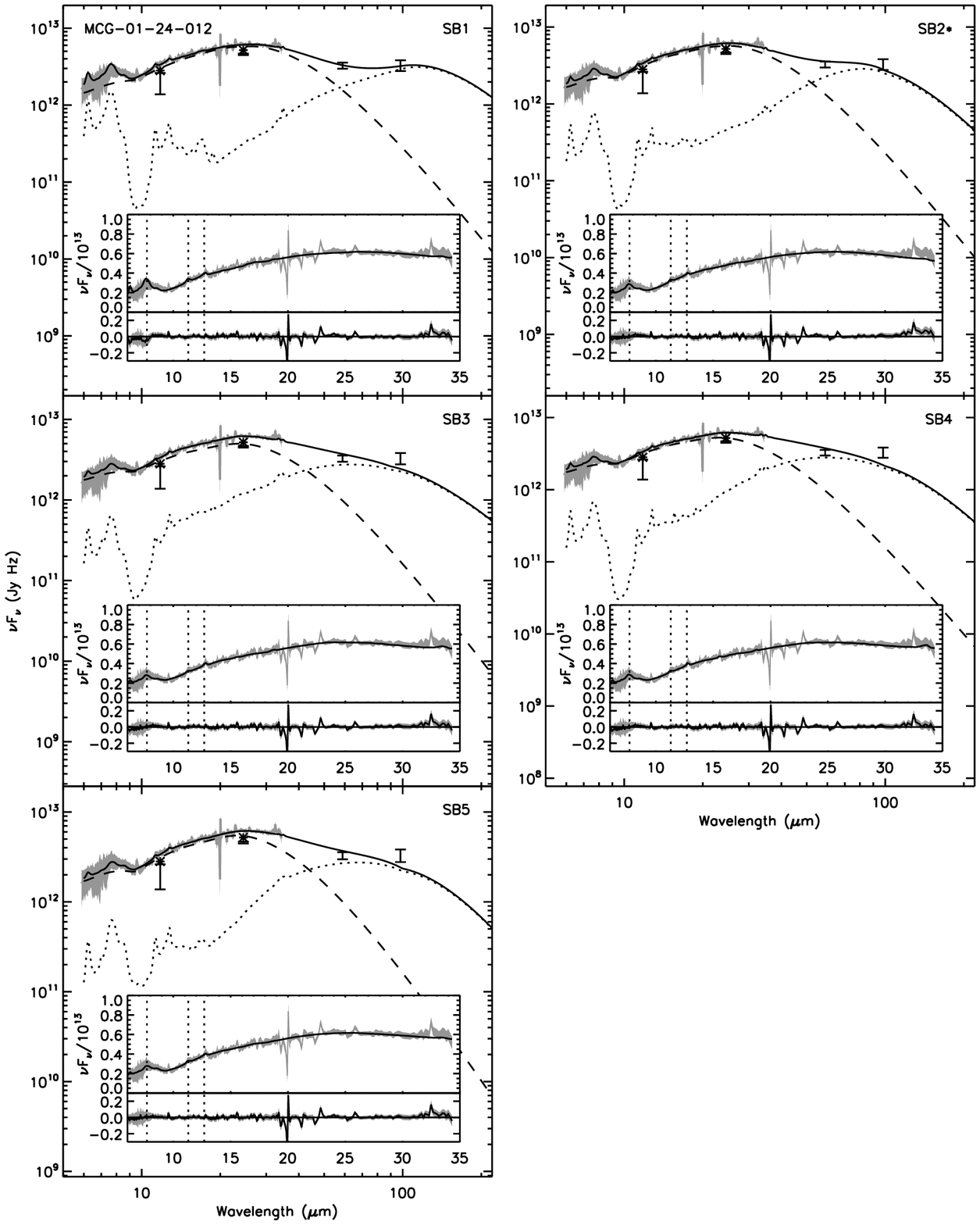}
\contcaption{}
\end{figure*}

\begin{figure*}
\includegraphics[width=160mm]{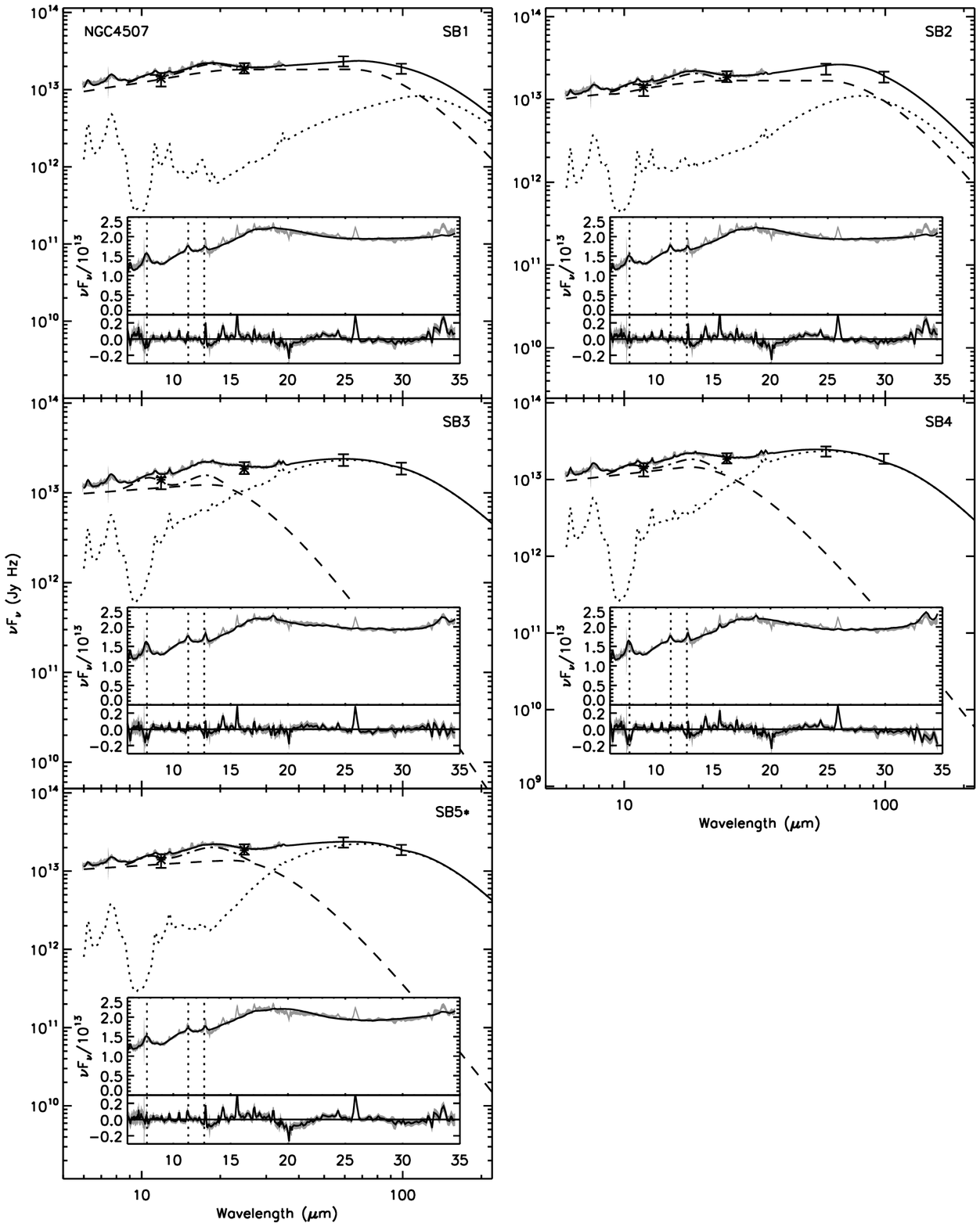}
\contcaption{}
\end{figure*}

\begin{figure*}
\includegraphics[width=160mm]{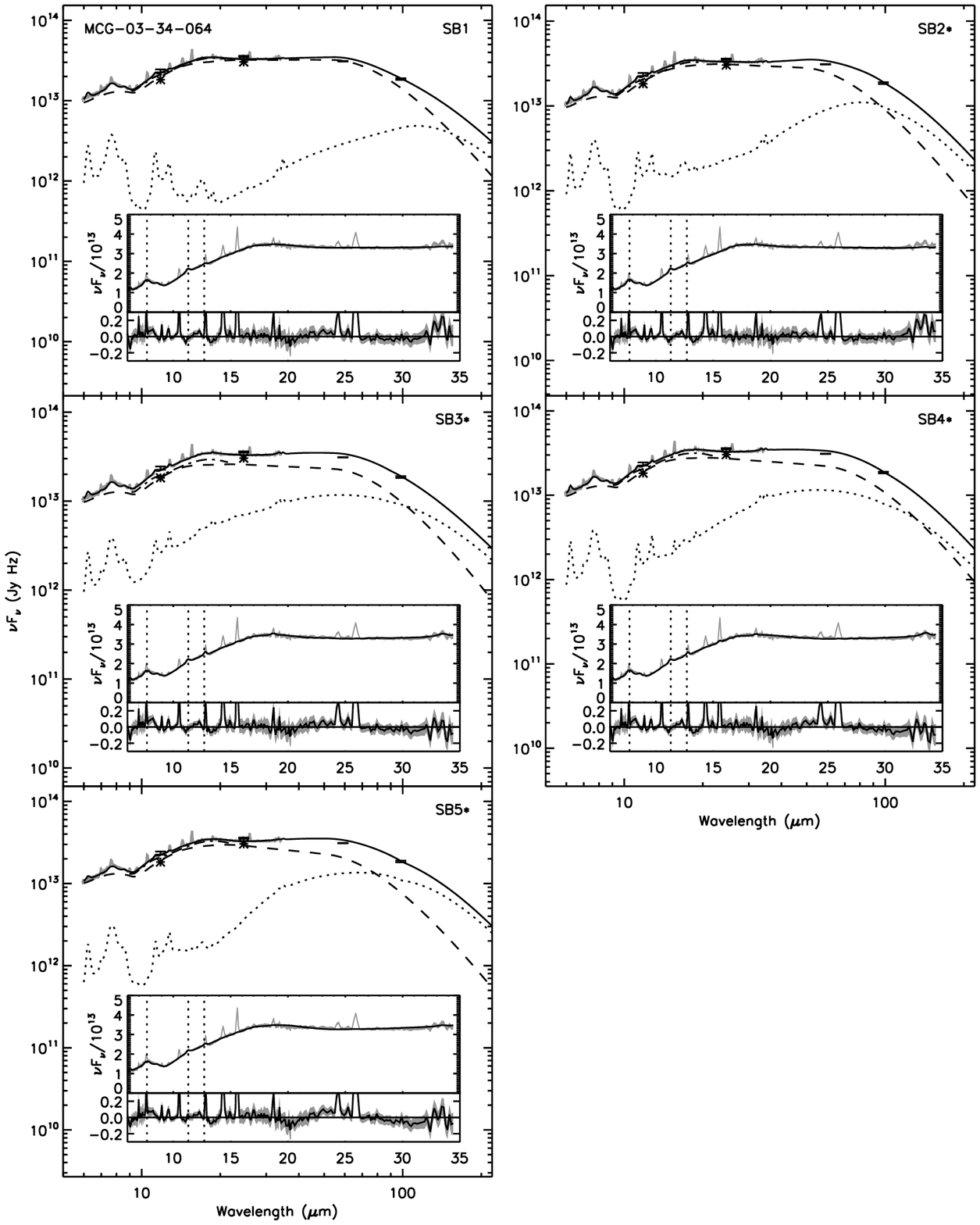}
\contcaption{}
\end{figure*}

\begin{figure*}
\includegraphics[width=160mm]{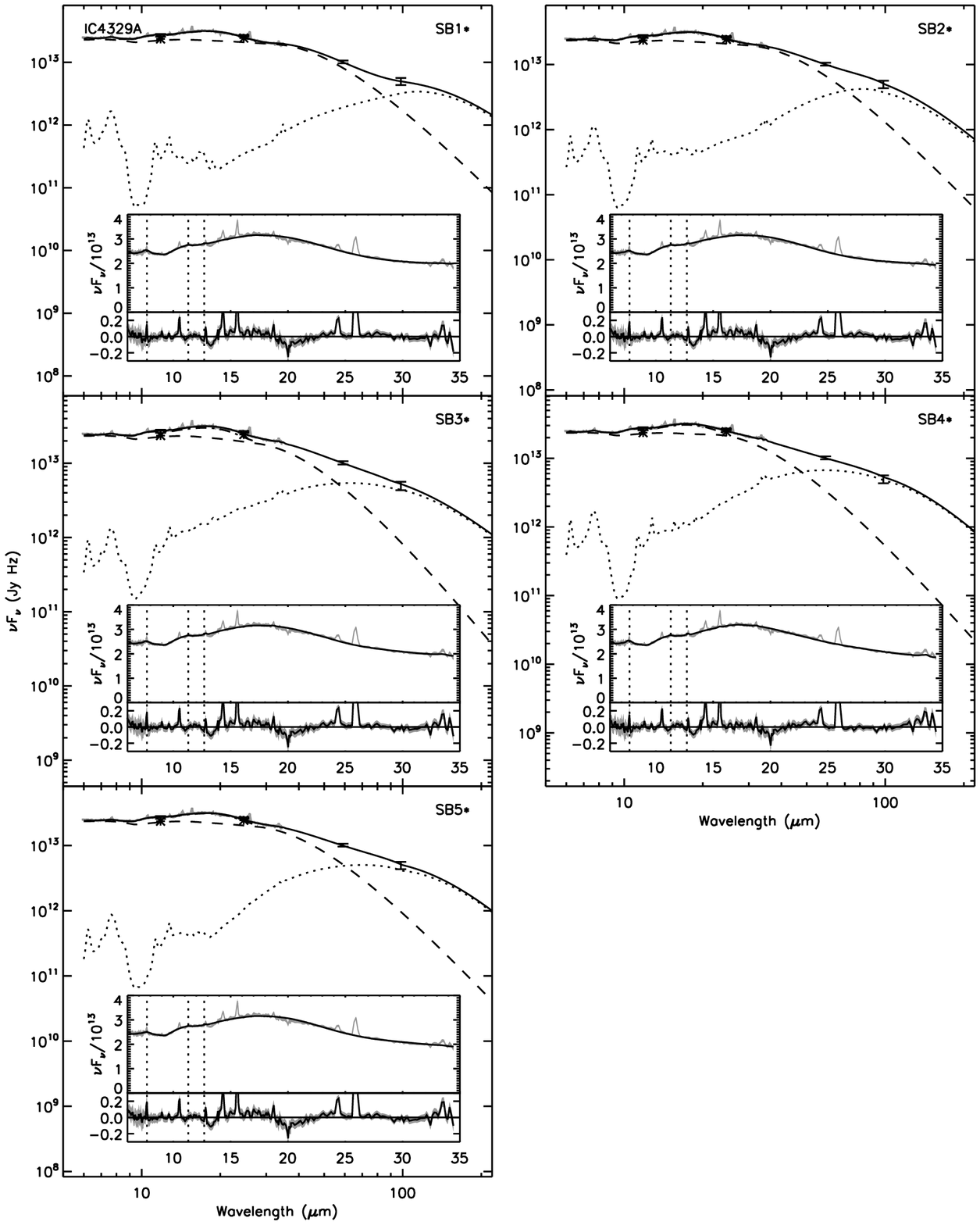}
\contcaption{}
\end{figure*}

\begin{figure*}
\includegraphics[width=160mm]{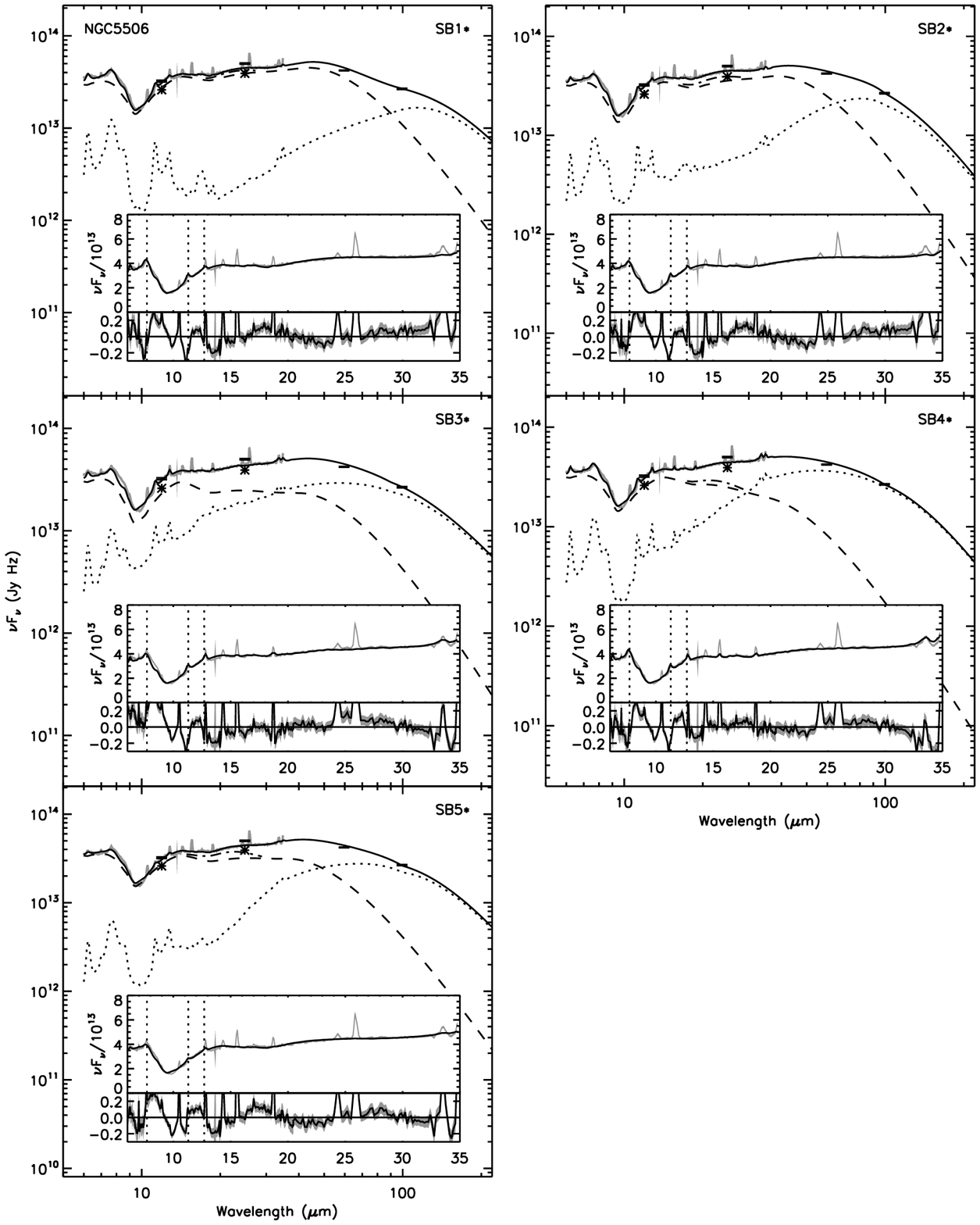}
\contcaption{}
\end{figure*}

\begin{figure*}
\includegraphics[width=160mm]{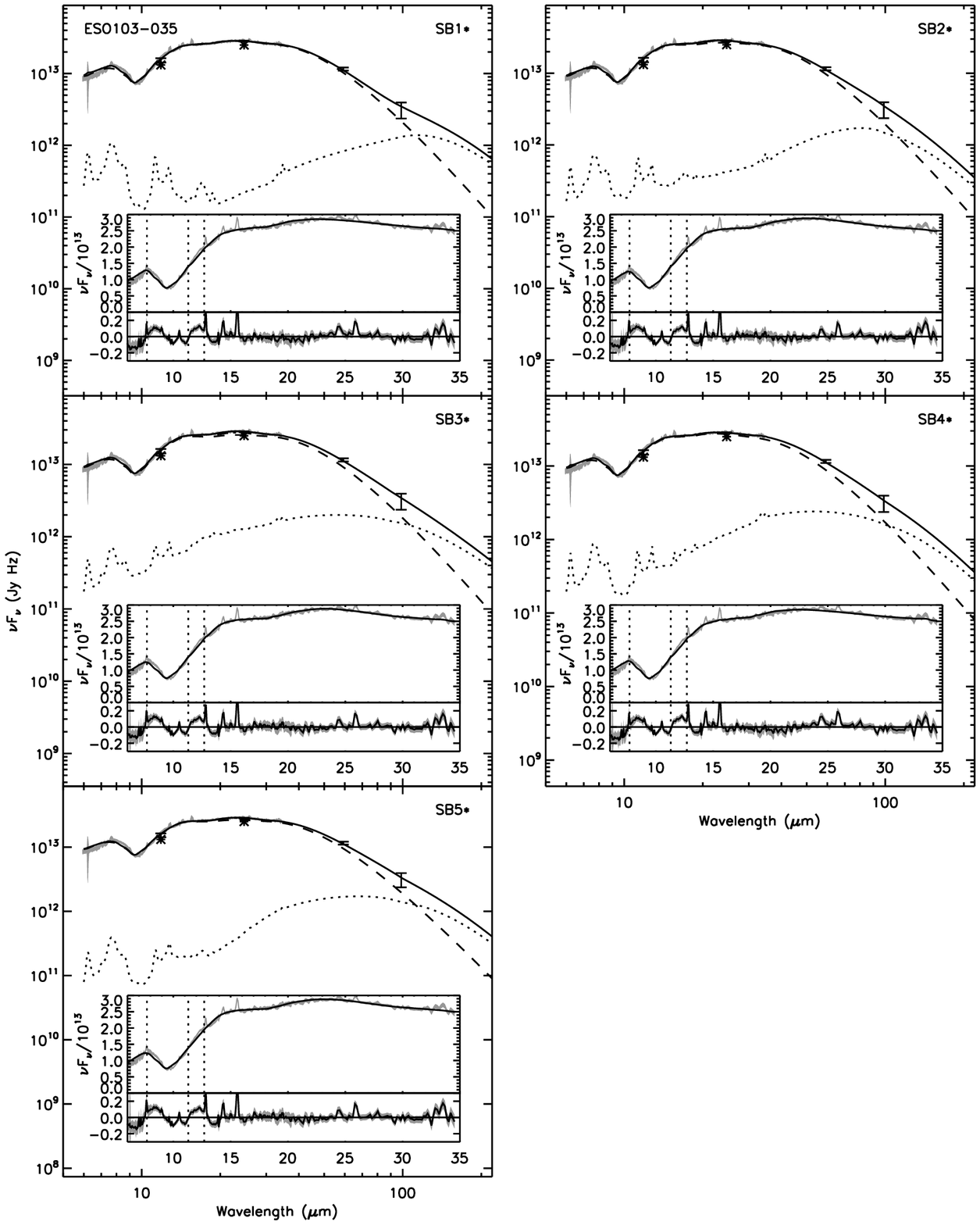}
\contcaption{}
\end{figure*}

\begin{figure*}
\includegraphics[width=160mm]{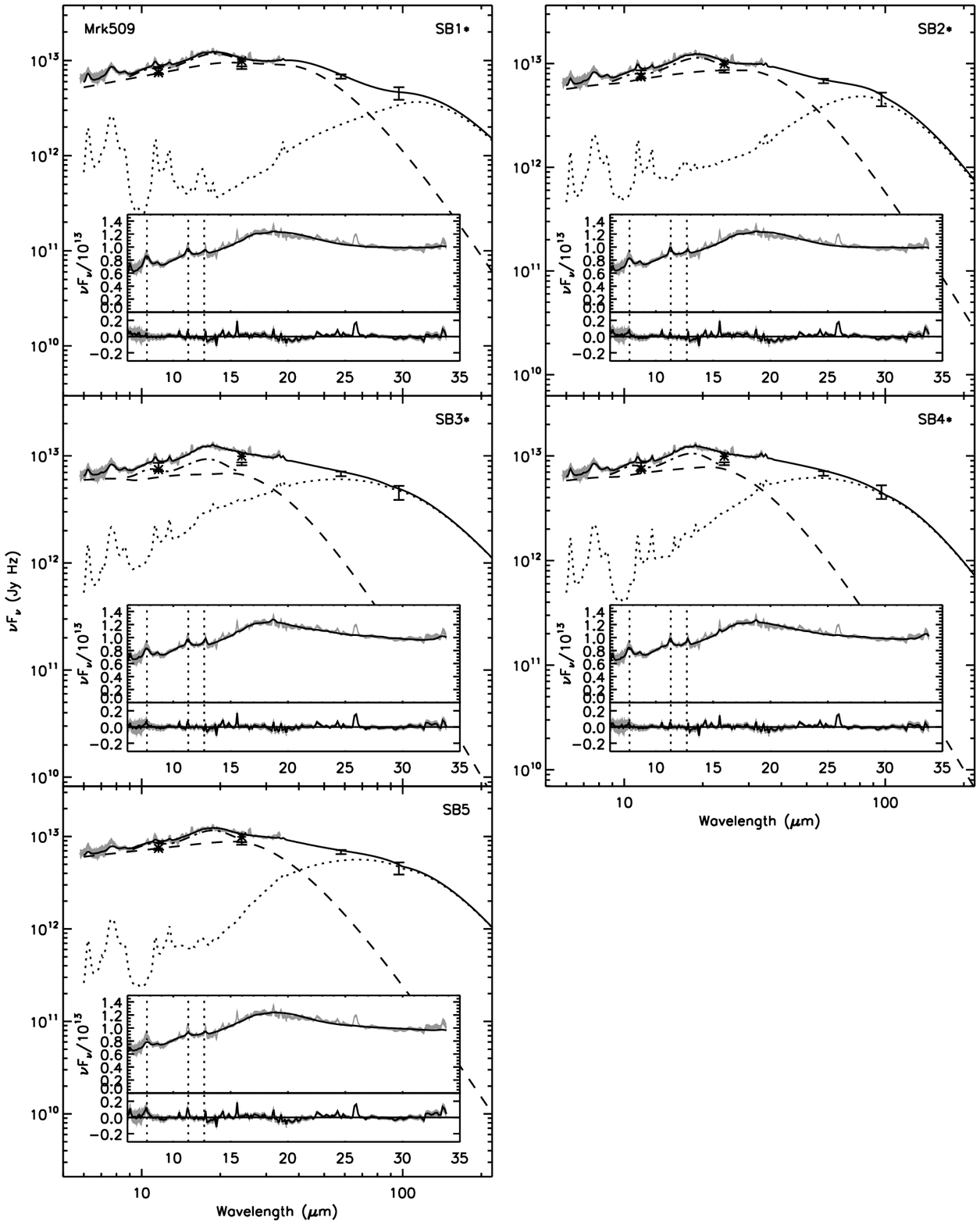}
\contcaption{}
\end{figure*}

\begin{figure*}
\includegraphics[width=160mm]{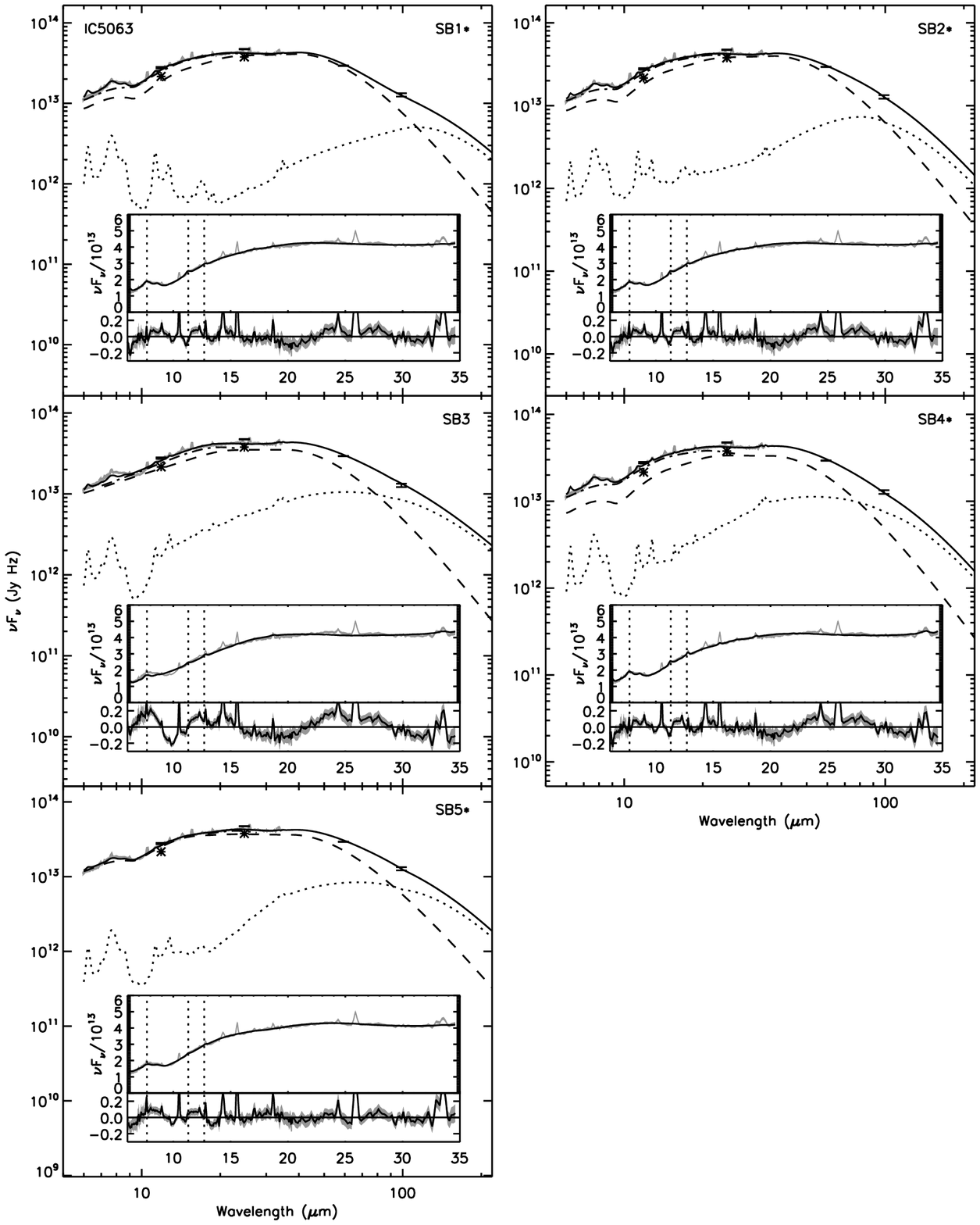}
\contcaption{}
\label{selectedAGNs2}
\end{figure*}

\section{Description of the fitting routine: {\sc decompir}} 
\label{Tests:Outline}

The results presented in \S\ref{Tests} and \S\ref{Application} were
obtained using a new infrared SED fitting routine, \dcmb, that we have
developed which uses the host-galaxy and intrinsic AGN SEDs described
in \S\ref{Templates} to measure the contribution to the infrared
output of composite galaxies from these two components.  In this
appendix, we provide a description of the main features of \dcmb\
routine and some brief instructions for its use.  \footnote{The {\sc
    idl} source code for \dcmb\ has been made available at
  http://sites.google.com/site/decompir}

\dcmb\ is written in the {\sc idl} programming language and uses
$\chi^2$ minimisation to identify the best fitting model SED to the
observed infrared data (either spectra and/or photometry) by varying
the values of a set of free parameters (e.g., component
normalisations, AGN component shape, level of dust extinction).
\dcmb\ makes heavy use of the {\sc idl} package {\sc mpfit} to perform
this fit.\footnote{{\sc mpfit} is also publicly available from
  http://purl.com/net/mpfit and is described in \protect
  \cite{Markwardt09}.}  In its default mode, \dcmb\ will attempt to
fit the supplied infrared photometry or spectrum with the average
intrinsic AGN infrared SED defined in \S\ref{Templates} plus
the ``SB1'' host galaxy template (see \S\ref{Templates:SB}).  By
default, only the normalisations of these components are allowed to
vary.  However, \dcmb\ can be instructed to allow any of the
parameters used to describe the AGN component to vary within the range
of values outlined in \S\ref{Templates} (i.e., $\alpha_1$,
$\alpha_2$, $\lambda_{\rm Brk}$, $\lambda_{\rm BB}$).  \dcmb\ can also
be instructed to allow the levels of extinction applied to both the
host-galaxy and AGN components to vary. We note that \dcmb\ will only
use one host-galaxy template at a time (i.e., an SED cannot be fit
with multiple host-galaxy components).  We therefore recommend that
each observed SED is fit five times, once for each of the host-galaxy
templates defined in \S\ref{Templates:SB}.  The best fitting SED can
then be selected from the five separate fits.

Because \dcmb\ relies on $\chi^2$ minimisation to derive the best
fitting parameters, the numbers of parameters that are allowed to vary
depends on the number of independent flux density measurements being
fit. Specifically, the number of variable parameters must be less than
the number of independent flux measurements to prevent the $\chi^2$
fits from becoming degenerate.  Some judgement based on the nature of
the research and the number of independent flux measurements is needed
in order to decide which parameters should be allowed to vary.  As a
guide, we provide a list of variable parameters in order of
importance, with the first having the largest effect on the fitted
SED: $\alpha_1$, dust extinction (both components), $\alpha_2$,
silicate emission features, $\lambda_{\rm BB}$, $\lambda_{\rm Brk}$
(we assume that the relative normalisations of the components are
always allowed to vary).  For example, if only five or six independent
flux measurements are available (e.g. \Spitzer\ 8~\mum, 24~\mum\ plus
\Herschel\ 100~\mum, 160~\mum, 250~\mum\ photometry) we suggest that
only $\alpha_1$, together with the normalisations of the AGN and
host-galaxy templates, is allowed to vary.  We note throughout the
main text that the choice of average SED template makes little
difference to the results from decomposing the photometric SEDs of
samples of composite galaxies.  The reason why this is the case is
that, at FIR wavelengths (i.e., $\lambda>30$~\mum) the SED is
typically (although not exclusively, see \S\ref{Templates:AGN:FIR})
dominated by the host-galaxy.  Therefore, a difference of a factor of
2-3 between the intrinsic AGN SED at 100~\mum\ has little effect on
our results when the observed SED is $\gtrsim 80$ per cent dominated
by the host galaxy at this wavelength.  However, extra care must be
taken in selecting the appropriate intrinsic AGN SED when it is
possible that the FIR SED is dominated by AGN emission (i.e., when the
observed SED falls strongly longwards of $\sim$30~\mum, e.g.,
MCG-03-34-064, IC5063, ESO103-035, Mrk~3, from this study; see
\fig{IntrinsicSEDs})

\label{lastpage}

\end{document}

%% file: table1.tex
\begin{tabular}{@{}lccccccccccccccc@{}}
\hline
\hline
Name&Type&RA&Dec.&Dist.&$S_{12}$&$S_{25}$&$S_{60}$&$S_{100}$&$S_{12}^{\rm IRS}$&$S_{25}^{\rm IRS}$&$\nu L_\nu$(60\micron)&\lir&\lx&\ewpah&Selected\\
&&&&(Mpc)&(Jy)&(Jy)&(Jy)&(Jy)&(Jy)&(Jy)&log(\lsun)&log(\lsun)&log(\ergs)&(\micron)&\\
(1)&(2)&(3)&(4)&(5)&(6)&(7)&(8)&(9)&(10)&(11)&(12)&(13)&(14)&(15)&(16)\\
\hline
Mrk348&2&12.1964&31.9570&63.4&0.308 $\pm$ 0.031&0.835 $\pm$ 0.025&1.29 $\pm$ 0.12&1.55 $\pm$ 0.20&0.1&0.4&9.91&10.5&42.6&0.002&-\\
NGC985&1&38.6574&-8.7876&185.3&0.207 $\pm$ 0.031&0.523 $\pm$ 0.017&1.38 $\pm$ 0.11&1.89 $\pm$ 0.15&0.1&0.3&10.87&11.3&44.3&0.016&-\\
NGC1275&2&49.9507&41.5117&74.2&1.060 $\pm$ 0.014&3.440 $\pm$ 0.027&6.990 $\pm$ 0.042&7.20 $\pm$ 0.47&0.7&2.8&10.78&11.2&43.9&-&-\\
ESO548-G081&1&55.5155&-21.2444&61.0&0.249 $\pm$ 0.057&0.097 $\pm$ 0.075&0.601 $\pm$ 0.048&1.49 $\pm$ 0.18&0.2&0.3&9.54&10.2&42.8&0.015&-\\
3C120&1&68.2962&5.3543&141.2&0.289 $\pm$ 0.043&0.624 $\pm$ 0.094&1.38 $\pm$ 0.21&1.94 $\pm$ 0.29&0.3&0.8&10.63&11.1&44.0&0.008&Y\\
NGC2110&2&88.0474&-7.4562&32.6&0.349 $\pm$ 0.028&0.840 $\pm$ 0.021&4.13 $\pm$ 0.21&5.7 $\pm$ 1.4&0.3&0.8&9.83&10.2&42.5&0.023&Y\\
EXO055620-3820.2&1&89.5083&-38.3346&144.9&0.529 $\pm$ 0.032&0.685 $\pm$ 0.034&0.322 $\pm$ 0.035&0.56 $\pm$ 0.12&0.3&0.6&10.02&11.2&43.7&-&-\\
Mrk3&2&93.9015&71.0375&56.9&0.713 $\pm$ 0.050&2.900 $\pm$ 0.036&3.77 $\pm$ 0.15&3.36 $\pm$ 0.44&0.6&2.3&10.28&10.8&42.9&-&Y\\
MCG-01-24-012&2&140.1927&-8.0561&83.1&0.080 $\pm$ 0.025&0.3810 $\pm$ 0.0088&0.654 $\pm$ 0.059&1.10 $\pm$ 0.18&0.1&0.4&9.85&10.4&43.0&0.006&Y\\
Mrk417&2&162.3789&22.9644&140.1&0.132 $\pm$ 0.033&0.227 $\pm$ 0.054&0.164 $\pm$ 0.043&0.71 $\pm$ 0.15&0.1&0.3&9.70&10.7&43.3&0.004&-\\
NGC3783&1&174.7572&-37.7386&38.5$^1$&0.840 $\pm$ 0.059&2.490 $\pm$ 0.050&3.26 $\pm$ 0.19&4.90 $\pm$ 0.54&0.6&1.9&9.88&10.5&43.1&0.002&-\\
NGC4507&2&188.9026&-39.9093&49.6&0.517 $\pm$ 0.078&1.59 $\pm$ 0.24&4.69 $\pm$ 0.70&6.28 $\pm$ 0.94&0.6&1.5&10.25&10.7&43.0&0.013&Y\\
ESO506-G027&2&189.7275&-27.3078&106.3&0.148 $\pm$ 0.030&0.268 $\pm$ 0.022&0.498 $\pm$ 0.050&0.83 $\pm$ 0.19&0.2&0.4&9.94&10.5&43.2&-&-\\
MCG-03-34-064&1.8&200.6019&-16.7286&69.8&0.940 $\pm$ 0.040&2.970 $\pm$ 0.045&6.200 $\pm$ 0.040&6.20 $\pm$ 0.14&0.7&2.5&10.67&11.1&42.8&0.005&Y\\
MCG-06-30-015&1.2&203.9741&-34.2956&32.5&0.380 $\pm$ 0.034&0.809 $\pm$ 0.027&1.090 $\pm$ 0.076&1.10 $\pm$ 0.22&0.2&0.5&9.25&9.9&42.6&0.004&-\\
IC4329A&1.2&207.3304&-30.3096&67.7&1.080 $\pm$ 0.054&2.210 $\pm$ 0.054&2.03 $\pm$ 0.10&1.66 $\pm$ 0.22&0.9&2.0&10.16&10.9&43.6&-&Y\\
NGC5506&1.9&213.3119&-3.2075&28.7$^1$&1.290 $\pm$ 0.027&4.170 $\pm$ 0.056&8.420 $\pm$ 0.060&8.87 $\pm$ 0.11&1.0&3.3&10.03&10.5&42.8&0.011&Y\\
NGC5548&1.5&214.4981&25.1368&72.5&0.401 $\pm$ 0.040&0.769 $\pm$ 0.032&1.070 $\pm$ 0.086&1.61 $\pm$ 0.16&0.2&0.6&9.94&10.6&43.5&0.017&-\\
Mrk290&1&233.9682&57.9026&126.1&$<$0.1&$<$0.1&0.171 $\pm$ 0.029&$<$0.6&0.1&0.2&9.63&$<$10.5&43.2&-&-\\
ESO103-035&2&279.5847&-65.4276&55.9&0.612 $\pm$ 0.043&2.360 $\pm$ 0.031&2.31 $\pm$ 0.12&1.05 $\pm$ 0.26&0.5&2.1&10.05&10.7&42.7&-&Y\\
3C390.3&1&280.5375&79.7714&244.3&0.128 $\pm$ 0.018&0.2870 $\pm$ 0.0089&0.204 $\pm$ 0.033&$<$0.6&0.1&0.3&10.28&$<$11.2&44.1&0.002&-\\
Mrk509&1.2&311.0406&-10.7235&147.3&0.316 $\pm$ 0.028&0.702 $\pm$ 0.022&1.360 $\pm$ 0.068&1.52 $\pm$ 0.23&0.3&0.8&10.66&11.2&44.0&0.023&Y\\
IC5063&2&313.0097&-57.0688&47.7&1.110 $\pm$ 0.023&3.940 $\pm$ 0.030&5.870 $\pm$ 0.038&4.25 $\pm$ 0.21&0.9&3.1&10.32&10.8&42.7&0.000&Y\\
NGC7213&1.5&332.3177&-47.1667&22.0$^1$&0.606 $\pm$ 0.049&0.742 $\pm$ 0.036&2.67 $\pm$ 0.16&8.18 $\pm$ 0.41&0.2&0.5&9.30&9.9&42.3&0.019&-\\
NGC7314&1.9&338.9426&-26.0502&19.0$^1$&0.268 $\pm$ 0.035&0.579 $\pm$ 0.048&3.74 $\pm$ 0.22&14.2 $\pm$ 1.3&0.1&0.5&9.32&9.8&42.2&0.015&-\\
\hline
\end{tabular}

%% file: table2.tex
\begin{tabular}{@{}lccc@{}}
\hline
\hline
Host Galaxy Template&Galaxy&D (Mpc)&log(\lir/\lsun)\\
(1)&(2)&(3)&(4)\\

\hline
SB1&NGC~1667&60.51&10.96\\
     &NGC~5734&59.28&11.06\\
     &NGC~6286&79.78&11.32\\
     &NGC~7590&21.58&10.16\\
SB2&NGC~7252&64.67$^a$&10.77$^c$\\
SB3&Mrk~52  &33.50$^b$&10.25$^c$\\
     &NGC~4818& 9.37&9.75\\
     &NGC~7714&38.16&10.72\\
SB4&NGC~1222&32.26&10.60\\ 
     &NGC~3256&35.35&11.56\\
     &NGC~4194&40.33&11.06\\
SB5&NGC~520 &30.22&10.91\\
     &NGC~660 &12.33&10.49\\
     &NGC~2623&77.43&11.54\\
\hline
\end{tabular}

%% file: table3.tex
\begin{tabular}{@{}lcccccccccc@{}}
\hline
\hline
&\multicolumn{3}{c}{$F_\nu^{\rm AGN}$}&&&&&\\
$\lambda$&Mean&Hi. Lum&Lo. Lum&$F_\nu^{\rm SB1}$&$F_\nu^{\rm SB2}$&$F_\nu^{\rm SB3}$&$F_\nu^{\rm SB4}$&$F_\nu^{\rm SB5}$&&\\
(\micron)&(Jy)&(Jy)&(Jy)&(Jy)&(Jy)&(Jy)&(Jy)&(Jy)&$\frac{L_{\rm IR}^{\rm AGN}}{\nu L_\nu^{\rm AGN}}$&$\frac{L_{\rm IR}^{\rm QSO}}{\nu L_\nu^{\rm QSO}}$\\
(1)&(2)&(3)&(4)&(5)&(6)&(7)&(8)&(9)&(10)&(11)\\
\hline
 6.0&  1.0&  1.0&  1.0&  1.0&  1.0&  1.0&  1.0&  1.0&  3.1&  1.9\\
 6.5&  1.1&  1.1&  1.1&  1.5&  1.3&  1.4&  1.4&  1.5&  3.0&  1.9\\
 7.0&  1.3&  1.2&  1.3&  1.9&  2.0&  2.1&  2.0&  2.4&  2.9&  2.0\\
 7.5&  1.4&  1.3&  1.4&  4.1&  4.3&  4.0&  4.3&  4.7&  2.8&  2.0\\
 8.0&  1.4&  1.5&  1.4&  3.3&  3.9&  3.6&  3.7&  4.5&  2.9&  1.9\\
 8.5&  1.4&  1.6&  1.4&  2.8&  3.7&  3.5&  3.4&  3.7&  3.1&  1.9\\
 9.0&  1.4&  1.7&  1.3&  1.1&  2.0&  2.7&  1.9&  1.8&  3.3&  1.7\\
 9.5&  1.5&  1.9&  1.3&  0.9&  1.8&  2.7&  1.5&  1.5&  3.3&  1.6\\
10.0&  1.7&  2.2&  1.5&  0.8&  1.8&  2.9&  1.5&  1.5&  3.0&  1.5\\
10.5&  2.0&  2.4&  1.8&  1.1&  2.3&  3.8&  2.2&  1.9&  2.7&  1.4\\
11.0&  2.3&  2.6&  2.1&  2.0&  3.3&  4.7&  3.0&  2.6&  2.6&  1.5\\
11.6&  2.5&  2.7&  2.5&  2.6&  4.5&  6.0&  4.2&  4.3&  2.4&  1.5\\
12.0&  2.8&  2.8&  2.8&  2.3&  4.3&  6.2&  4.3&  4.7&  2.3&  1.6\\
12.6&  3.0&  2.9&  3.1&  3.2&  5.6&  7.8&  5.9&  6.7&  2.2&  1.7\\
\hline
\end{tabular}